# OFDM Reference Signal Pattern Design Criteria for Integrated Communication and Sensing

Rui Zhang, Shawn Tsai, Tzu-Han Chou, Jiaying Ren, Wenze Qu and Oliver Sun

*Abstract*— Extended ambiguity performance (EAP), which includes all grating lobes and side peaks, indicates the maximum detectable region without undesired peaks for target parameter estimation and is critical to radar sensor design. Driven by EAP requirements of bi-static sensing, we propose design criteria for orthogonal frequency division multiplexing (OFDM) reference signal (RS) patterns. The design not only improves EAP in both time delay and Doppler shift domains under different types of sensing algorithms, but also reduces resource overhead for integrated communication and sensing. With minimal modifications of post-FFT processing for current RS patterns, guard interval is extended beyond conventional cyclic prefix (CP), while maintaining inter-symbol-interference-(ISI)-free delay estimation. For standard-resolution sensing algorithms, a staggering offset of a linear slope that is relatively prime to the RS comb size is suggested. As for super-resolution sensing algorithms, necessary and sufficient conditions of comb RS staggering offsets, plus new patterns synthesized therefrom, are derived for the corresponding achievable EAP. Furthermore, we generalize the RS pattern design criterion for super-resolution sensing algorithms to irregular forms, which minimizes number of resource elements (REs) for associated algorithms to eliminate all side peaks. Starting from staggered comb pattern in current positioning RS, our generalized design eventually removes any regular form for ultimate flexibility. Overall, the proposed techniques are promising to extend the ISI- and ambiguity-free range of distance and speed estimates for radar sensing.

## I. INTRODUCTION

IN recent years there is a growing interest in integrated communication and sensing (ICAS) for the 5th generation advanced (5G-A) and the upcoming 6th generation (6G) mobile communication systems [1-4]. It is envisioned that wireless networks will evolve to add sensing functions to support emerging intelligent applications such as activity or object detection, recognition, and tracking [5-7]. The communication system and sensing system can share waveform, reference signals, and most hardware components, and mutually assist each other. One of the research directions is communication centric ICAS, which realizes sensing functionality in a system primarily designed for communications [4, 5, 8, 9]. In such a system, it is preferred for sensing function to reuse current wireless communication signals such as orthogonal frequency division multiplexing (OFDM) adopted by 5G New Radio (NR) [8, 10, 11]. There have been extensive research efforts of utilizing the OFDM signal defined in cellular and Wi-Fi networks for radar sensing based on various sensing algorithms such as matched filter, maximum likelihood detection, estimation of signal parameters via rotational invariance technique (ESPRIT) [12-14], etc.

For bi-static sensing, where ICAS data payload is not known at the receiver side, communication RS is the most straightforward radio resource for the sensing receiver to estimate time delays and Doppler shifts caused by targets [3, 15]. Mono-static sensing can leverage both the RS and payload [16]. On the other hand, power budget in mono-static sensing is critical due to the power loss is dependent on round-trip propagation. In pathloss dominating scenarios, reference signal has the potential to achieve better PAPR performance than conventional OFDM signal due to special sequence designs. The time delays and the Doppler shifts are proportional to sensing targets' ranges and radial velocities, respectively. Communication RS also possesses similar features desired by sensing signal such as good auto-correlation properties and low peak-to-average power ratio (PAPR) [17]. Thus, there has been research on reusing communication RS for sensing, e.g., Long-Term Evolution (LTE) RSs and positioning reference signal (PRS) for radar sensing in [18-20]. A joint reference signal design and power optimization is proposed for energy-efficient 5G vehicle-to-everything (V2X) ICAS [21]. There are various types of RS in 5G NR system: demodulation RS (DMRS), channel state information RS (CSI-RS), tracking RS (TRS), phase tracking RS (PTRS), and PRS [17]. Most communication RSs were designed for channel state information and communication receiver demodulation. Among them, 5G PRS has a special design of multiple comb RS symbols with a cycle of staggering offsets, which is specifically designed for delay estimation but not for both delay and Doppler estimation. Therefore, extra aspects derived from general sensing criteria require another look at other possibilities. Specifically, with Doppler shift added to the estimated variables, ambiguities and side peaks arise in the two-dimensional (2D) delay-and-Doppler plane. Existing work mostly focused on legacy reference signals or dedicated sensing waveforms. In this pa-per, further generalization of OFDM RS design, including different time and frequency densities, staggering offsets, and irregular RS patterns, is first attempted for improved sensing performance and resolution with small overheads. In radar sensing, resolutions and ambiguities-plus-side-peaks of both delay and Doppler are two important criteria for system design. Super-resolution sensing algorithms can also improve sensing resolution but require higher computational complexity than standard-resolution algorithms. The standard resolution algorithms considered here include Delay-and-Sum and periodogram (such as 2D FFT) [22-24]. In [25], we studied the staggered comb reference signal design based on aforementioned standard-resolution algorithms. In this paper, we will extend our study to super-resolution algorithms and irregular reference signal design. As for super-resolution sensing algorithms, we use two common ones in radar sensing, iterative adaptive approach (IAA) and 2D Multiple Signal Classification (MUSIC), as examples [26-33]. Fundamentally the total bandwidth and the time duration of a sensing signal determines the achievable resolutions for delay and Doppler, respectively. Nonetheless, in the next section, we will show that tuning the comb RS resource elements (REs) density in time and frequency not only adjusts the resolutions but also changes ambiguities and side peaks in both delay and Doppler domains. Hereafter ambiguities and side peaks caused by signal format and receiver algorithm are referred to as extended ambiguity performance (EAP). And being free from both ambiguities and undesired side peaks will be specified as unambiguous.

Our contributions of OFDM RS patterns design are summarized as follows.
1) Based on standard-resolution sensing algorithms, we derive the design principles of staggering offsets over multiple symbols that could help improve EAP in both the delay and the Doppler domains while maintaining a good resolution performance, compared with state-of-art RS.
2) We introduce an extended guard interval design for comb RS to enable ISI removal beyond CP, which helps extend the sensing range of algorithms performed after FFT and descrambling.
3) We propose a general rule of designing staggering offsets over multiple comb RS symbols and their RE spacing in the 2D time and frequency domain, when a super-resolution type of sensing algorithm is used. The design achieves the best EAP in 2D time delay and Doppler shift domain with system overhead lower than what is required by standard resolution algorithms.
4) We derive necessary and sufficient conditions of synthesizing two different comb RS patterns to reach the best achievable EAP when super-resolution algorithms are used. The design lowers system overhead compared to 3).
5) We derive necessary and sufficient conditions of generalized irregular RS patterns to achieve the best EAP for super-resolution algorithms. The design lowers system overhead compared to 3).

The paper is organized as follows. First, we will review comb RS patterns from current wireless RS in Section II, which will be used for analyses in following sections. Then we will define and use Delay-and-Sum algorithm as an example to introduce the criteria of improving 2D EAP in Section III. Different comb RS patterns using Delay-and-Sum and their performance characterized by 2D-unambigous range (defined therein) will be investigated. Next in Section IV, we propose the extended guard interval design to extend the sensing distance for post-FFT algorithms such as 2D FFT, IAA, MUSIC, etc. Afterwards, we will analyze the EAP of different comb RS patterns using 2D FFT in Section V. Then we will derive the design criteria of comb RS patterns for super-resolution algorithms using IAA as an example in Section VI. We will synthesize different comb RSs into new patterns in Section VII and generalize the analysis to novel irregular RS pattern for super-resolution algorithms in Section VIII. Finally, a conclusion section summarizes the paper.

## II. GENERAL COMB RS PATTERNS WITH UNIFORM SYMBOL SPACING

A general comb RS pattern can be characterized by Fig. 1 with the following parameters:
1) $S_{sub}$ (unit in subcarrier numbers) is the RS frequency domain spacing of REs and $S_{sub} \geq 2$ for comb RS. $S_{sub}$ is also called comb size, which determines the frequency domain RE density,
2) $S_{sym}$ (unit in symbol numbers) is the time domain separation of the RS symbols (RE density in the time domain),
3) $F_i$ (unit in subcarrier numbers) is the frequency domain RE offset (or the staggering offset) of the $i$-th RS symbol, and
4) $M$ and $N$ are the number of sensing RS symbols and total number of subcarriers, respectively. The total time span from the first sensing RS to the last sensing RS is $MS_{sym}T$, where $T$ is time duration of a CP-OFDM symbol.

One can tune the RS pattern by $S_{sub}$, $S_{sym}$, $F_i$, $M$ and $N$. Fig. 1 shows one example with $S_{sub} = 4$, $S_{sym} = 2$, $(F_0, F_1) = (0,2)$, $M = 2$, and $N = 12$.

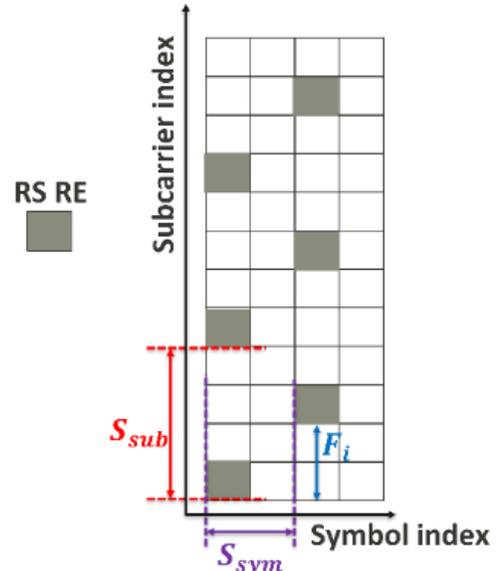

**Fig. 1.** Illustration of general comb RS pattern.

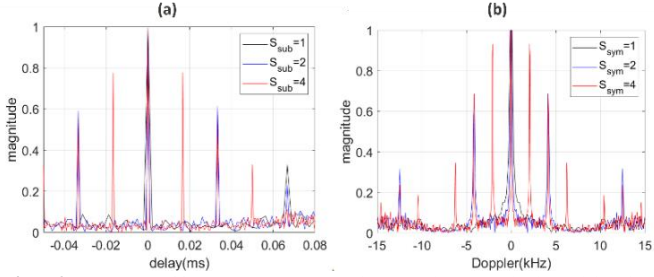

**Fig. 2:** Trade-off between resolution and extended ambiguity.

The comb RS has several benefits. First, it boosts energy per RE for better coverage. Second, it maximizes the delay resolution with smaller number of REs spanning across the entire channel bandwidth, compared to subband RE-block allocation. Third, it enables either different base-station PRS multiplexing or data RE insertion. However, the comb RS introduces delay ambiguity at a fractional symbol level for standard resolution algorithms. Therefore, the comb RS pattern over multiple symbols with a cycle of staggering offsets is adopted in 5G PRS to preserve up-to-one-effective-OFDM-symbol unambiguous delay [17].

Nonetheless, sensing requires another look at other RS pattern possibilities. Specifically, with Doppler frequency shift added to the estimated variables, undesired side peaks arise in the 2D delay-and-Doppler plane. Under a fixed number of REs, larger $S_{sub}$ and $S_{sym}$ increase occupied bandwidth and time duration, which enhance range (time delay) and velocity (Doppler shift) resolutions, respectively. However, such resolution improvement by tuning $S_{sub}$ and $S_{sym}$ comes with costs of reduced maximum unambiguous delay and Doppler shift. The delay resolution is proportional to $\frac{1}{\Delta f K_1 S_{sub}}$, and the maximum unambiguous delay is proportional to $\frac{1}{\Delta f S_{sub}}$, where $\Delta f$ is subcarrier spacing, $K_1$ is number of RE in the frequency domain. As for Doppler estimation, the Doppler resolution is proportional to $\frac{1}{MTS_{sym}}$ and the maximin unambiguous Doppler is proportional to $\frac{1}{TS_{sym}}$. Fig. 2 illustrates that increasing $S_{sym}$ and $S_{sub}$ respectively enhance Doppler and delay resolution as the 3dB mainlobe width decreases. However, side peaks occur closer to the mainlobe and result in worse EAP although the resolution performance improves. In the following section III to VI, we will analyze how to design staggering offsets of different RS patterns such that one can still maintain a good EAP while increasing $S_{sym}$ ($S_{sub}$).

Several RS patterns with uniform symbol spacing are shown in Fig. 3 and categorized by Scheme A, B, C and D for the remaining part of the paper:

(a) Fig. 3(a) illustrates an example of Scheme A with $S_{sub} = 4, S_{sym} = 1$, which is similar to 5G PDCCH DMRS or TRS. This pattern keeps the same RE locations across all RS symbols, i.e., $F_i$ stays constant for any $i$-th RS symbol.

(b) Fig. 3(b) depicts an example of Scheme B, which is similar to partial PRS (subgroup of a full staggering cycle) with $S_{sub} = 4, S_{sym} = 1, F_0 = 0$, and $F_1 = S_{sub}/2$,. In 5G NR, partial PRS has the format of staggering two RS symbols' REs by half an even comb size, (i.e., for $F_0 = f, F_1 = f + S_{sub}/2$).

(c) Fig. 3(c) shows an instance of Scheme C with $S_{sub} = 4$, $S_{sym} = 1$, using the 5G PRS staggering offset sequence. When $S_{sym} = 1$, the RS pattern is a 5G PRS.

(d) Fig. 3(d) illustrates an instance of the proposed Scheme D with $S_{sub} = 4, S_{sym} = 1$, and $p = 1$, which is linear-slope staggering with the slope relatively prime to the comb size. The staggering offset sequence is $F_i = \mod(p \cdot (i-1), S_{sub})$ for $i = 0, \cdots, S_{sub} - 1$, where $p$ is relatively prime to $S_{sub}$.

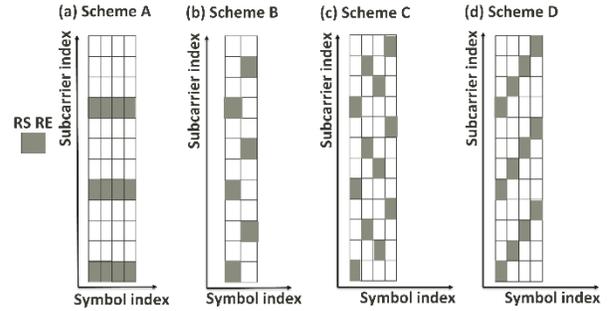

**Fig. 3.** Instances of RS patterns.

## III. AMBIGUITY PERFORMANCE OF DELAY-AND-SUM

The ambiguity function (AF) of a time-domain signal $s(t)$ based on Delay-and-Sum is expressed as [22]

$$A(\tau, f) = \frac{1}{E_s}\left|\int_{-\infty}^{\infty} s(t)e^{(j2\pi ft)}s^*(t-\tau)dt\right|, \quad (1)$$

where $E_s = \int_{-\infty}^{\infty} s(t)s^*(t)dt$, $f$ is the Doppler frequency shift, and $\tau$ is the time delay. Note that the ambiguity function is for delay and sum. For other sensing algorithms, we will analyze their ambiguity performance using EAP. Moreover, super-resolution algorithms could help suppress extended ambiguities [34]. We will show later that super-resolution algorithms could achieve better EAP with a proper design of RS. In general, the design criterion is to minimize $A(\tau, f)$ in the areas where $(\tau, f) \neq (\tau^*, f^*)$, where $\tau^*$ and $f^*$ are the nominal true time delay and Doppler. To simplify the description in the following sections, $(\tau^*, f^*)$ is set to (0,0) without loss of generality (WLOG) as AF is a form of RS pattern's impulse response in the 2D domain. Large values of $A(\tau, f)$ exceeding a radar dynamic range at $(\tau, f) \neq (0, 0)$ are undesired side peaks. These side peaks limit the unambiguously detectable range of time delay and Doppler frequency shift. When there are side peaks, the design criteria then become how to enlarge the 2D maximum unambiguous region, or how to design the waveform such that side peak locations $(\tau, f)$ are as far away from (0, 0) as possible. For instance, in radar sensing, Costas array has been designed based on such criteria [35, 36]. The design criteria also apply to 2D FFT and super-resolution sensing algorithms.

We now derive a general expression of AF based on Delay-and-Sum and the general comb RS patterns in Section II. Let $T_s$ be the effective OFDM symbol duration (the reciprocal of the subcarrier spacing (SCS)), $T_{cp}$ be the CP duration, $T = T_{cp} + T_s$ be the CP-added OFDM symbol duration, and $X_i$ be the frequency-domain RE scrambling sequence of the $i$-th comb RS. The $i$-th RS symbol sample vector of length $N$ over duration $T_s$ is denoted as $Y = \{Y(n)\}_{n=0}^{N-1}$, where the $n$-th sample is

$$Y(n) = \sum_{\kappa=0}^{N/S_{sub}-1} X_i(\kappa S_{sub} + F_i) e^{\frac{j2\pi(\kappa S_{sub}+F_i)n}{N}}, \quad (2)$$

Note that $N$ is the number of subcarriers and $N/S_{sub}$ denotes the number of REs of comb sensing RS. Observe that

$$Y(n' + \frac{Nl}{S_{sub}}) = Y(n')e^{\frac{j2\pi F_i l}{S_{sub}}}, \quad (3)$$

where $n' = 0,1,2,\cdots,N/S_{sub} - 1$ and $l = 0,1,2,\cdots,S_{sub} - 1$. Thus $Y$ can be further divided equally into $S_{sub}$ subsets, where the $l$-th subset is denoted by $Y_l$, each of length-$(N/S_{sub})$ and

$$Y_l = Y_0 e^{j\left(\frac{2\pi F_i l}{S_{sub}}\right)}, Y_l = \{Y(n)\}_{n=lN/S_{sub}}^{(l+1)N/S_{sub}-1}. \quad (4)$$

Let the time sequence of CP-added OFDM symbol be $Z = \{Z(n)\}_{n=0}^{N'}$, and the $n$-th sample of $Z$ over the CP-added duration $T$ be

$$Z(n) = \sum_{\kappa=0}^{N/S_{sub}-1} X_i(\kappa S_{sub} + F_i) e^{\frac{j2\pi(\kappa S_{sub}+F_i)(n-N_{cp})}{N}}, \quad (5)$$

where $N' = (N + N_{cp})$. Let $M$ be the number of RS symbols, the AF by Delay-and-Sum over $M$ RS symbols is expressed by:

$$A(\tau,f) = \left| \sum_{i=0}^{M-1} \sum_{n=iS_{sym}N'+\frac{Nl}{S_{sub}}}^{iS_{sym}N'+N'-1} Z(n)Z^*\left(n - \frac{\tau N}{T_s}\right) e^{\frac{j2\pi fnT}{N}} \right|. \quad (6)$$

As time samples of one OFDM symbol possess $S_{sub}$ subsets that are repetitive with piecewise phase rotation as shown by Eq. (4), strong side peaks of a Delay-and-Sum receiver happen at $\tau = l \cdot T_s/S_{sub} = l \cdot N/S_{sub} \cdot (T_s/N)$, where $l \cdot N/S_{sub}$ is the beginning sample index of the $l$-th subset of Eq. (3). In this case, the design requires a sequence with good auto correlation property such that the AF at $\tau \neq l \cdot T_s/S_{sub}$ has negligible values. Then we can focus on the analysis of AF by plugging $\tau = l \cdot T_s/S_{sub}$ into Eq (6). Assuming $X_i$ preserves constant envelope in the time domain (e.g., frequency-domain Zadoff-Chu sequence), we obtain the AF at time delay $\tau = l \cdot T_s/S_{sub}$ and Doppler frequency $f$ over $M$ RS symbols as:

$$A\left(\frac{T_s l}{S_{sub}}, f\right) = \left| S \cdot e^{\frac{j2\pi fIT}{S_{sub}}} \sum_{i=0}^{M-1} e^{-\frac{j2\pi F_i l}{S_{sub}}} \cdot \sum_{n=iS_{sym}N'}^{iS_{sym}N'+N'-1-\frac{Nl}{S_{sub}}} e^{\frac{j2\pi fnT}{N}} \right|, \quad (7)$$

where $S = Z(n)Z^*(n)$ is a constant. Doppler frequency is detected by a variable-$f$-phase-rotated version of the time sequence. In the case of $f = 0$ (e.g., positioning), the staggering offset sequence $\{F_i\}$ directly affects the side peak locations by

$$A\left(\frac{T_s l}{S_{sub}}, 0\right) = \left| S \cdot \left(N' - \frac{Nl}{S_{sub}}\right) \sum_{i=0}^{M-1} e^{-\frac{j2\pi F_i l}{S_{sub}}} \right|, \quad (8)$$

For instance, $M = S_{sub}$, $\{F_i\}_{i=0}^{S_{sub}-1} = \{0,1,\cdots,S_{sub}-1\}$, the side peaks along the axis of $f = 0$ in the time delay range $(0, T_s)$ can be eliminated as $\sum_{i=0}^{M-1} e^{-\frac{j2\pi F_i l}{S_{sub}}} = 0$ for all $l \in \{1,\cdots,S_{sub}-1\}$. AF characteristics derived from Delay-and-Sum algorithm over various RS patterns will be examined next in this section.

We will use 2D unambiguous range to characterize the maximum Doppler and maximum time delay that sensing algorithms can detect without any ambiguities in the analysis. We name it 2D maximum unambiguity hereafter. The red dotted rectangle in Fig. 4(b) is an example. The 2D unambiguous range is determined by the associate side peaks of the mainlobe. For any targets with a mainlobe in the 2D unambiguous range, their associated side peaks fall outside the red rectangle, meaning no ambiguities exist in the designated 2D range. Therefore, the RS design needs to enlarge the 2D unambiguous range with an efficient utilization of REs. The ambiguity function is simulated under the setting of 15kHz SCS.

*Scheme A: Patterns similar to 5G PDCCH DMRS or TRS*

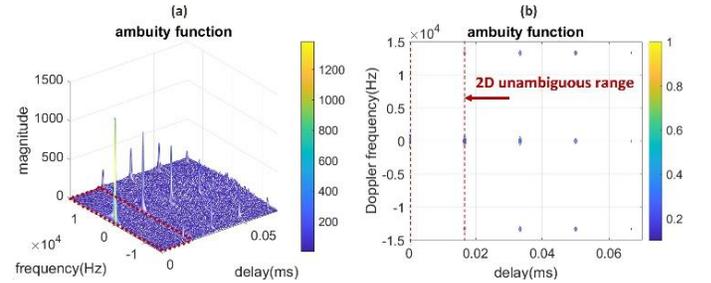

Fig. 4. AF in the case of $(S_{sub} = 4, S_{sym} = 1, F_i = F_j = 0)$.

Based on Eq. (7) and Eq. (8), Scheme A ($F_i \equiv$ constant without staggering) has 2D AF side peak locations at $(\tau = lT_s/S_{sub}, f = k/S_{sym}T)$, where $l, k \in \mathbb{Z}$ (set of integers), $0 \leq l \leq S_{sub}, -S_{sym} \leq k < S_{sym}$ and $(l, k) \neq (0,0)$. Note that there is an exception for Delay-and-Sum to be free of ambiguities at $(\tau = 0, f = \pm 1/T)$ because there is no discrete-time Fourier transform to create frequency domain aliasing. If the RS symbols do not have time interruption (i.e., $S_{sym} = 1$), the maximum unambiguous Doppler frequency for Delay-and-Sum is only limited by the sampling rate $N'/T$ because there is no frequency-domain convolution with a non-impulse. Therefore, the range of unambiguous Doppler is $[-1/2S_{sym}T, 1/2S_{sym}T]$ for $S_{sym} > 1$ and $[-N'/2T, N'/2T]$ for $S_{sym} = 1$. Those side peaks creating ambiguities for Delay-and-Sum results are shown by the example in Fig. 4 with $S_{sub} = 4$, $S_{sym} = 1$, $F_i \equiv 0$, under SCS=15 kHz, where Fig. 4(b) is the top-down view of Fig. 4(a). Fig. 5 shows another example with $S_{sub} = 8$, $S_{sym} = 2$, and $F_i \equiv 0$. Note that the increased comb size ($S_{sub}$ from 4 to 8) reduces the maximum unambiguous time delay, and the interruption of RS continuation in time ($S_{sym} = 2$ instead of 1) causes convolution with a gating function's frequency-domain response, creating ambiguities along the true delay's Doppler frequency axis before reaching $N'/T$. However, the maximum unambiguous delay in this case is always restricted to $T_s/S_{sub}$.

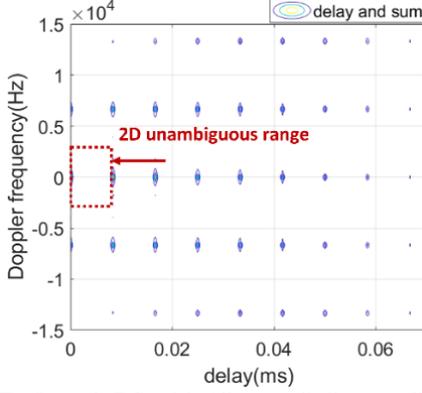

**Fig. 5.** AF of comb RS with $(S_{sub} = 8, S_{sym} = 2, F_i \equiv 0)$.

*Scheme B: Patterns similar to partial PRS*

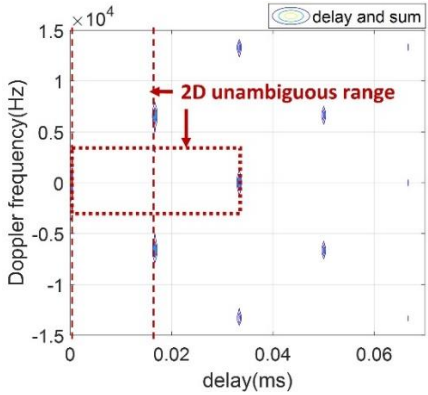

**Fig. 6.** AF of Partial PRS with $(S_{sub} = 4, S_{sym} = 1, F_0 = 0, F_1 = 2)$.

In the case of Scheme B, half-comb-size staggering is incorporated, creating alternating $S_{sub}$-subset ambiguities. That is, for even integers $l_1$, side peaks occur at $(\tau = l_1 T_s/S_{sub}, f = k_1/S_{sym}T)$, for $0 \le l_1 \le S_{sub}$, $-S_{sym} \le k_1 \le S_{sym}$, $k_1 \in \mathbb{Z}$, and $(l_1, k_1) \ne (0,0)$, where the aforementioned $(\tau = 0, f = \pm 1/T)$ exception still holds. For odd integers $l_2$, side peaks occur at locations $(\tau = l_2 T_s/S_{sub}, f = (1/2 + k_2)/(S_{sym}T))$, for $0 \le l_2 \le S_{sub}, -S_{sym} \le k_2 \le S_{sym}$, $k_2 \in \mathbb{Z}$ ) and $(l_2, k_2) \ne (0,0)$. Fig. 6 shows the half-comb-size-staggering example of $F_1 = 0$, and $F_2 = 2$ under $S_{sub} = 4$ and $S_{sym} = 1$,. Such RS pattern increases the maximum unambiguous delay from $T_s/S_{sub}$ to $2 \cdot (T_s/S_{sub})$ compared with Scheme A. Depending on application scenarios, the designated 2-D maximum unambiguity, beside the vertically long dotted rectangle in Fig. 6 as analyzed before, can be opted for an increased maximum unambiguous delay from 0 to $2T_s/S_{sub}$, and a Doppler range now from $-1/(4S_{sym}T)$ to $1/(4S_{sym}T)$. Such a new option is shown by the horizontally wider dotted red rectangle of Fig. 6.

*Scheme C: Patterns with staggering offset of 5G PRS*

Fig. 7 shows an AF of Scheme C with $S_{sub} = 4, S_{sym} = 1$ and $\{F_0, F_1, F_2, F_3\} = \{0, 2, 1, 3\}$, which is a staggering offset sequence such that the REs sweep through the whole signal bandwidth. This PRS pattern can increase the maximum unambiguous time delay to full-symbol length $T_s$, albeit at the cost of smaller Doppler unambiguity. Based on Eq. (8), the horizontally wider dotted red rectangle of Fig. 7 shows an option to designate full-symbol delay unambiguity if the application scenario precludes large Doppler, for which $f$ is assumed to be between $-1/(2S_{sym}S_{sub}T)$ and $1/(2S_{sym}S_{sub}T)$.

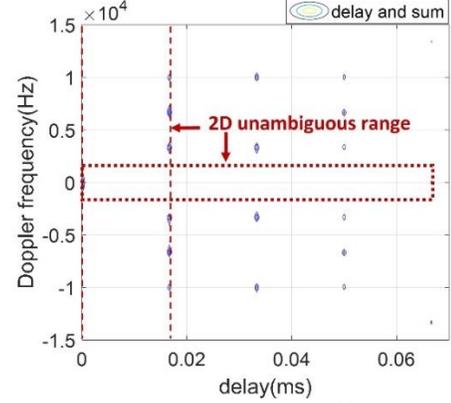

**Fig. 7.** AF of PRS-like pattern with $(S_{sub} = 4, S_{sym} = 1, \{F_0, F_1, F_2, F_3\} = \{0,2,1,3\}$.

*Scheme D: Linear-slope-relatively-prime-to-comb-size staggering*

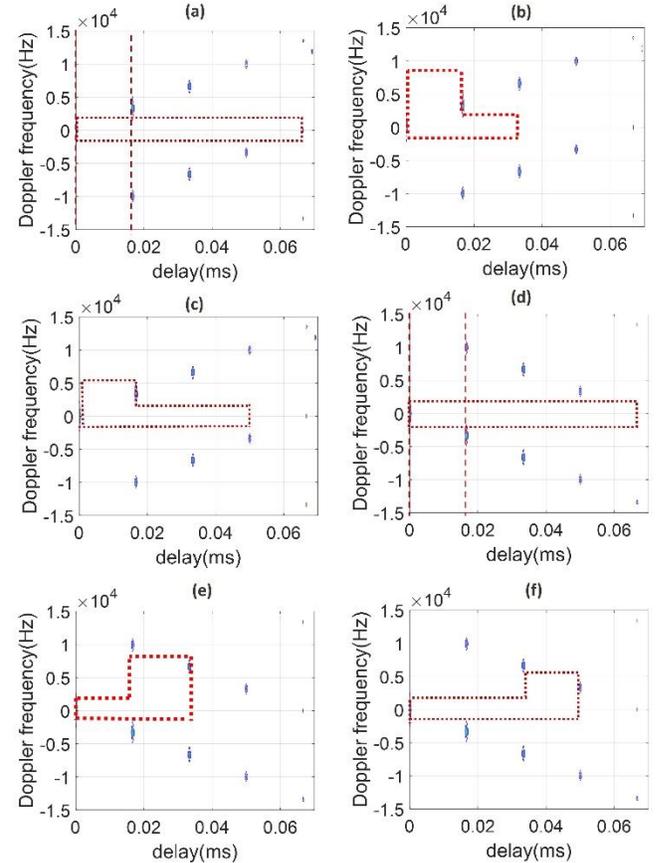

Fig. 8. AF in the case of $(S_{sub} = 4, S_{sym} = 1)$, proposed staggering Scheme D, where in (a), (b) and (c) $p = 1$, and in (d), (e) and (f) $p = 3$.

Now we turn to the proposed Scheme D, where the staggering offset sequence $\{F_i\}_{i=0}^{S_{sub}-1}$ has a linear slope $p$ that is relatively prime to the comb size $S_{sub}$. Corresponding side peaks occur at locations ($\tau = l \cdot T_s/S_{sub}$, $f = p \cdot l / S_{sub}S_{sym}T + k/S_{sym}T$) for $0 \le l \le S_{sub}$, $|k| \le S_{sym}$, $(l,k) \ne (0,0)$ and the exception at $(0, \pm 1/T)$. Compared with those of aforementioned RS patterns, the proposed scheme shows more flexibility in determining 2D maximum unambiguity. The AF of $S_{sub} = 4$, $S_{sym} = 1$, and $p = 1$ is shown in Fig. 8(a) and (b) with different choices of designating unambiguous regions. Fig. 8(c) and (d) present the AF of $S_{sub} = 4$, $S_{sym} = 1$, and $p = 3$ (a different slope that is relatively prime to $S_{sub}$). In summary, the staggering offset sequence $F_i = \mod(p \cdot (i-1), S_{sub})$, where $p$ is relatively prime to $S_{sub}$, is suggested for Delay-and-Sum sensing algorithms as it provides more flexible 2D maximum unambiguity choices for time delay and Doppler frequency shift detection.

## IV. EXTENDED GUARD INTERVAL FOR POST FFT SENSING ALGORITHMS

Post-FFT frequency domain sensing algorithms, e.g., 2D FFT [23, 24] as well as super-resolution algorithms [24-31], are conventionally limited by CP duration, which determines the maximum time delay (hence distance) without ISI for the OFDM ICAS system. In this section, we will show that the comb RS pattern with zero-power REs in between could extend the guard interval against ISI beyond CP. Flexibility of extending ISI-free time delay up to CP plus $(S_{sub}-1)/S_{sub}$ of effective OFDM symbol length can thus be achieved by post-FFT algorithms.

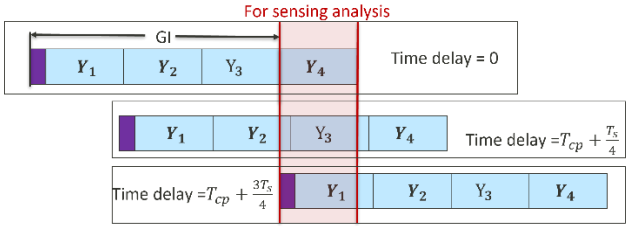

Fig. 9. Sensing analysis based on $(1/S_{sub})$-symbol duration.

Assuming the Doppler shift is sufficiently small compared to the OFDM base frequency, i.e., $|f| < f_{max} \ll 1/T_s$ (where $f_{max}$ is typically 1/10 of SCS), we can approximate it as a constant phase rotation block-by-block over each OFDM symbol. Such an assumption was not required in the case of Delay-and-Sum over a continuous period for estimating the Doppler frequency. However, for post-FFT sensing algorithms, it is required as time observations of Doppler frequency has been discretized at the granularity of an OFDM symbol. Recall CP-added OFDM time samples as defined in Eq. (5), an extended guard interval against sensing ISI (GI in Fig. 9) is now defined as the first $(N_{cp} + lN/S_{sub})$ samples, where $l \in \{0, 1, \cdots, S_{sub}-1\}$) denotes the fractional symbol index. Such extension can reuse the same CP-OFDM symbol (originally designed for shorter multipath delay communication) to longer distance sensing without resorting to a new waveform.

Let the number of sensing targets in the channel be $H$, the time delay and Doppler frequency introduced by the $h$-th target be $\tau_h$ and $f_h$, respectively, and $\alpha_h$ be the complex amplitude coefficient of each target path. After removing the extended guard interval against ISI, the remaining time samples of the $i$-th OFDM symbol, are expressed by the time sample vector of the last $(S_{sub} - l)$ subsets $\boldsymbol{G}_l = \{G_l(m)\}$, for $m = 0, 1, \cdots, N(S_{sub} - l)/S_{sub} - 1$. The $m$-th element of $\boldsymbol{G}_l$ is

$$G_l(m) = \sum_{h=0}^{H-1} \alpha_h Z\left(\frac{Nl}{S_{sub}} + N_{cp} + m - \frac{\tau_h}{T_s}N\right) \cdot e^{j2\pi f_h iTS_{sym}}$$

$$= \sum_{h=0}^{H-1} \sum_{\kappa=0}^{\frac{N}{S_{sub}}-1} \alpha_h X_i(\kappa S_{sub} + F_i) e^{j\frac{2\pi(\kappa S_{sub}+F_i)\left(\frac{Nl}{S_{sub}}+m-\frac{\tau_h}{T_s}N\right)}{N}} e^{j2\pi f_h iTS_{sym}}.$$

(9)

where $0 \le N_\tau \le \min(N_{cp} + Nl/S_{sub}, N)$. To recover the non-zero REs from $\boldsymbol{G}_l$, the receiver side should first perform the following phase de-rotation procedure before FFT, that is

$$G'(m) = G_l(m) e^{\frac{-j2\pi F_i\left(\frac{Nl}{S_{sub}}+m\right)}{N}} \quad (10)$$

for $m = 0, 1, \cdots, N(S_{sub}-l)/S_{sub} - 1$. Then, after FFT of $\boldsymbol{G}'$, we obtain sequence $\boldsymbol{B}$ with a length of $N \cdot (S_{sub}-l)/S_{sub}$ and

$$B(\kappa') = \frac{1}{N}\sum_{m=0}^{\frac{(S_{sub}-l)N}{S_{sub}}-1} G'(m) e^{j\frac{-2\pi\kappa'S_{sub}m}{(S_{sub}-l)N}}. \quad (11)$$

In Eq. (11), for $\kappa' \in \{(S_{sub}-l)w, w = 0, 1, \cdots, N/S_{sub}-1\}$, we can write

$$B(\kappa') = \sum_{h=0}^{H-1} \alpha_h \frac{S_{sub}-l}{S_{sub}} \cdot X_i\left(\frac{\kappa'S_{sub}}{S_{sub}-l}+F_i\right) \cdot e^{j2\pi\left(f_h iTS_{sym} - \frac{\tau_h}{T_s}\left(\frac{\kappa'}{S_{sub}-l}S_{sub}+F_i\right)\right)}.$$

(12)

As for $\kappa' \notin (S_{sub}-l)w$, then $B(\kappa') = 0$. The frequency-domain sequence of the $i$-th RS symbol for sensing algorithm analysis (with a length of $N - lN/S_{sub}$) can be re-assembled to sequence $\boldsymbol{X}'_i$ with the original length of $N$. For $B(\kappa') \ne 0$, we then map $X'_i(\kappa'S_{sub}/(S_{sub}-1) + F_i) = B(\kappa')$. For RE locations other than $\kappa'S_{sub}/(S_{sub}-1) + F_i$, $\boldsymbol{X}'_i$ will be padded with zeros. Therefore, $\boldsymbol{X}'_i$ is further simplified as

$$X'_i(\kappa_1) = \sum_{h=0}^{H-1} \alpha_h \frac{(S_{sub}-l)}{S_{sub}} \cdot X_i(\kappa_1) e^{j2\pi f_h iTS_{sym}} e^{-\frac{2\pi\tau_h\kappa_1}{T_s}},$$

(13)

where $\kappa_1 = 0, 1, \ldots, N-1$. Thus, the maximum time delay without ISI is $\min(T_{cp} + lT_s/S_{sub}, T_s)$, parameterized by the fractional symbol index $l$. Note that there is a trade-off between signal energy loss and maximum time delay with different choices of $l$. Larger $l$ can support larger ISI-free time delay estimation but yields lower signal energy for sensing algorithms analysis. Fig. 9 illustrates an instance of $l = S_{sub} - 1$, and $S_{sub} = 4$.

## V. EAP OF 2D FFT

For 2D FFT, the conclusion is similar to Delay-and-Sum, with the exception that side peaks appear at $(0, \pm 1/T)$ due to frequency-domain aliasing brough by discrete-time Fourier transform. The advantage is that 2D FFT does not require special time sequence since the post-FFT sensing algorithm can descramble received signals by frequency domain scalar division without decorrelation. In addition to independence from scrambling sequence properties, 2D FFT also eliminates side peaks through staggering comb patterns, similarly to what Delay-and-Sum does. Based on Eq. (13), the 2D FFT algorithm performs a second FFT operation to a sequence of $M$ descrambled post-FFT OFDM symbols $\{X_{R,i}\}_{i=0}^{M-1}$, where the $\kappa_1$-th element of $X_{R,i}$ is:

$$X_{R,i}(\kappa_1) = \begin{cases} \dfrac{S_{sub}X'_i(\kappa_1)}{(S_{sub}-l)X_i(\kappa_1)} = \sum_{h=0}^{H-1} \alpha_h e^{j2\pi f_h i T S_{sym}} e^{-j2\pi \frac{\tau_h \kappa_1}{T_s}}, & X'_i(\kappa_1) \neq 0 \\ 0, & \text{otherwise} \end{cases} \quad (14)$$

for $\kappa_1 = 0, 1 \ldots N-1$. WLOG, we derive the 2D FFT result with the number of targets $H$ setting to one. The 2D FFT EAP is then written as:

$$P(g,q) = \left| \sum_{\kappa_1=0}^{N-1} \left( \sum_{i=0}^{M-1} X_{R,i}(\kappa_1) e^{-j2\pi \frac{qiS_{sym}}{M}} \right) e^{j2\pi \frac{g\kappa_1}{N}} \right|^2$$

$$= \left| \sum_{k=0}^{\frac{N}{S_{sub}}-1} \beta_1(g,k) \cdot \sum_{i=0}^{M-1} \beta_2(g,q,i) \right|^2 \quad (15)$$

For the first summation in the above equation, where $\beta_1(g,k) = e^{-j2\pi \frac{\tau_0 k S_{sub}}{T_s}} e^{j2\pi \frac{gk S_{sub}}{N}}$, the maximum value is obtained at $\beta_1(g,k) = 1$. That is, for

$$\frac{gS_{sub}}{N} - \frac{\tau_0 S_{sub}}{T_s} = z_1 \in \mathbb{Z}, \quad (16)$$

there will be side peaks in the 2D FFT result at the discrete-valued time delay domain, namely $gT_s/N = \tau$-axis. The side peaks appear at $\tau = \frac{gT_s}{N} = \tau_0 + \frac{z_1 T_s}{S_{sub}}$. Substituting $g = \frac{Nz_1}{S_{sub}} + \frac{\tau_0 N}{T_s}$, terms of the second summation where $\beta_2(g,q,i) = e^{j2\pi f_0 i T S_{sym}} e^{-j2\pi \frac{qiS_{sym}}{M}} e^{-j2\pi \frac{\tau_0}{T_s} F_i} e^{j2\pi \frac{gF_i}{N}}$ of Eq. (15) become

$$\sum_{i=0}^{M-1} \beta_2(g,q,i) = \sum_{i=0}^{M-1} e^{j2\pi f_0 i T S_{sym}} e^{-j2\pi \frac{qiS_{sym}}{M}} e^{j2\pi \frac{z_1 F_i}{S_{sub}}}. \quad (17)$$

Eq. (17) is the Fourier transform of $e^{j2\pi \frac{z_1 F_i}{S_{sub}}}$ evaluated at $f_0 T S_{sym} - \frac{qS_{sym}}{M}$, and thus different choices of $F_i$ yield different ambiguity and side peaks.

Taking the example of $F_i$ being constant over $i$, terms of the summation in Eq. (17) will reach the maximum when

$$\frac{q}{MT} = f_0 - \frac{z_2}{S_{sym}T}, \quad (18)$$

that is, $e^{j2\pi f_0 i T S_{sym}} e^{-j2\pi \frac{qiS_{sym}}{M}} = 1$. The delay and Doppler extended ambiguities in $\frac{q}{MT} = f$-axis (Doppler domain) and $\tau$-axis (delay domain) are $\left( f_0 - \frac{z_2}{S_{sym}T} \right)$ and $\left( \frac{z_1 T_s}{S_{sub}} + \tau_0 \right)$, respectively, where $z_1 \neq 0$, $z_2 \neq 0$ and $\tau_0 = \frac{N_\tau T_s}{N}$. WLOG, we assume the true time delay and Doppler frequency pair as $(\tau_0, f_0) = (0,0)$ in the analysis.

Fig. 10(a) shows the EAP of $S_{sub} = 4$, $S_{sym} = 1$, and $F_i \equiv 0$ (Scheme A). Similar to Delay-and-Sum, the maximum unambiguous delay is restricted to $T_s/S_{sub}$. Compared with Delay-and-Sum, the maximum unambiguous Doppler is smaller due to side peaks at $(0, \pm 1/T)$. Moreover, when $F_i$ is not linearly changing with $i$ (e.g., 5G PRS), the Fourier transform generates several side peaks and results can be calculated numerically. Fig. 10(b) and (c) show 2D maximum unambiguity in the case of $S_{sub} = 4$ and $S_{sym} = 1$ for Scheme B and C, respectively.

When staggering offset Scheme D is adopted, the Fourier transform of Eq. (17) yields delta functions in the Doppler domain at $\tau = \tau_0 + \frac{z_1 T_s}{S_{sub}}$. Substituting $F_i = p \cdot i$ into terms of Eq. (17), it will reach the maximum when $e^{j2\pi f_0 i T S_{sym}} e^{-j2\pi \frac{qiS_{sym}}{M}} e^{j2\pi \frac{z_1 pi}{S_{sub}}} = 1$, that is, $f_0 T S_{sym} - \frac{qS_{sym}}{M} + \frac{z_1 p}{S_{sub}} = z_2 \in \mathbb{Z}$, where $p$ is relatively prime to $S_{sub}$. This can be further written as

$$\frac{q}{MT} = f_0 - \frac{z_2}{S_{sym}T} + \frac{z_1 p}{S_{sub} S_{sym}T}. \quad (19)$$

Thus, delay and Doppler ambiguities in $\tau$-axis and $f$-axis are $MT\left( f_0 - \frac{z_2}{S_{sym}T} + \frac{z_1 p}{S_{sub} S_{sym}T} \right)$ and $\left( \frac{z_1 T_s}{S_{sub}} + \tau_0 \right)$, respectively, where $z_2 \neq 0$, $\tau_0 = \frac{N_\tau T_s}{N}$. Fig. 10(d) to (f) presents different choices of 2D maximum unambiguity in the case of $S_{sub} = 4$, $S_{sym} = 1$, and $(\tau_0, f_0) = (0,0)$ with the proposed Scheme D. Similarly, Scheme D is also suggested for 2D FFT due to flexibility in tuning 2D maximum unambiguity.

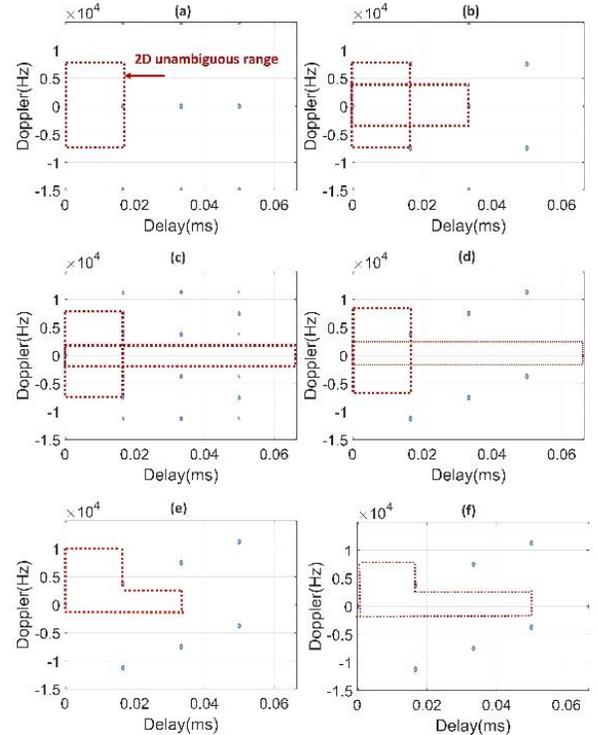

**Fig. 10**. EAF of 2D FFT with different RS patterns.

## VI. EAP OF COMB RS USING SUPER-RESOLUTION ALGORITHMS

To simultaneously detect the target velocity and range, here we leverage 2D MUSIC and 2D IAA as examples of super-resolution sensing algorithms. Based on Eq. (14) in previous section, after FFT and descrambling, let the received RS signal matrix of $M$ RS symbols in one snapshot be

$$\mathbf{F} = \begin{pmatrix} X_{R,0}(0) & \cdots & X_{R,M-1}(0) \\ \vdots & \ddots & \vdots \\ X_{R,0}(N-1) & \cdots & X_{R,M-1}(N-1) \end{pmatrix}. \quad (20)$$

For 2D MUSIC or 2D IAA [26-29], in one snapshot, we vectorize the matrix $\mathbf{F}$ to $\mathbf{A} = vec(\mathbf{F})$. Let $\mathbf{S}$ be the vector of complex-valued coefficients of targets, and $\mathbf{W}$ be the steering matrix we want to estimate (which does not change over snapshots), we write $\mathbf{A} = \mathbf{W}\mathbf{S} + \mathbf{N}'$, where $\mathbf{N}'$ is the vectorized noise, and $\mathbf{S} = [\alpha_0 \ \cdots \ \alpha_{H-1}]^T$ is the vector of targets' complex coefficients. The $h$-th column of the steering matrix $\mathbf{W}$ can be written as:

$$\mathbf{w}_{h1}(f_h, \tau_h) = \begin{bmatrix} e^{-\frac{j2\pi(0 \cdot S_{sub}+F_0)\tau_h}{T_s}} e^{j2\pi 0 \cdot S_{sym}Tf_h} \\ e^{-\frac{j2\pi(S_{sub}+F_0)\tau_h}{T_s}} e^{j2\pi 0 \cdot S_{sym}Tf_h} \\ \vdots \\ e^{-\frac{j2\pi((N-1) \cdot S_{sub}+F_0)\tau_h}{T_s}} e^{j2\pi 0 \cdot S_{sym}Tf_h} \\ \vdots \\ e^{-\frac{j2\pi(0 \cdot S_{sub}+F_{M-1})\tau_h}{T_s}} e^{j2\pi(M-1) \cdot S_{sym}Tf_h} \\ e^{-\frac{j2\pi(S_{sub}+F_{M-1})\tau_h}{T_s}} e^{j2\pi(M-1) \cdot S_{sym}Tf_h} \\ \vdots \\ e^{-\frac{j2\pi((N-1)S_{sub}+F_{M-1})\tau_h}{T_s}} e^{j2\pi(M-1) \cdot S_{sym}Tf_h} \end{bmatrix}. \quad (21)$$

The ambiguities due to aliasing happen when $\mathbf{w}_{h1}(f, \tau) = \mathbf{w}_{h1}(f', \tau')$ and $(f, \tau) \neq (f', \tau')$. We will discuss several cases of RS pattern and their EAP next.

### 1) Scheme A: Patterns similar to 5G PDCCH DMRS or TRS

If $F_i$ is constant for any $i$-th RS symbol, ambiguities happen when

$$e^{-j\left(\frac{2\pi(k \cdot S_{sub})}{T_s}\tau'\right)} e^{j2\pi i \cdot S_{sym}Tf'} = e^{-j\left(\frac{2\pi(k \cdot S_{sub})}{T_s}\tau\right)} e^{j2\pi i \cdot S_{sym}Tf}, \quad (22)$$

for all $k = 0,1,\cdots,N-1$, $i = 0,1,\cdots,M-1$. From Eq. (22), the condition is equivalent to $\tau' = \tau + k_1 T_s/S_{sub}$ and $f' = f + k_2/(S_{sym}T)$, where $\tau$ and $f$ are true delay and Doppler, respectively, and $(k_1, k_2)$ are arbitrary integer pairs.

Fig. 11(a) presents the EAP of 2D IAA with (0,0) as the true delay and Doppler pair in the case of $S_{sym} = 1, S_{sub} = 4$, under 15kHz SCS without staggering. The maximum unambiguous delay is always restricted to $T_s/S_{sub}$. Fig. 11(b) shows the EAP of $S_{sym} = 2, S_{sub} = 4$. Overall, the side peak's locations are the same as 2D FFT but with a better resolution.

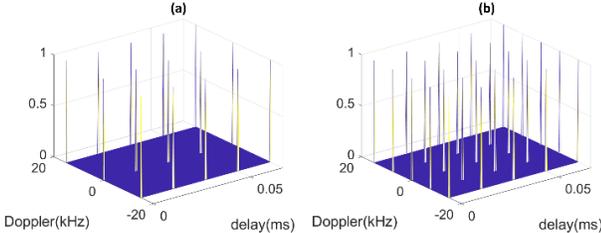

**Fig. 11.** Ambiguity performance with no staggering offsets using IAA.

### 2) Scheme D: Linear-slope-relatively-prime-to-comb-size staggering

Staggering Scheme D is defined as a linear-slope staggering offset such that $F_i = \mathrm{mod}(p \cdot i + \beta_1, S_{sub})$, where $p$ is relatively prime to $S_{sub}$ and $\beta_1 \in \{0,1,\cdots,S_{sub}-1\}$, for $i = 0,\cdots,M-1$. The ambiguities happen when

$$e^{-j\left(\frac{2\pi(k \cdot S_{sub}+ip)}{T_s}\tau'\right)} e^{j2\pi i \cdot S_{sym}Tf'} = e^{-j\left(\frac{2\pi(k \cdot S_{sub}+ip)}{T_s}\tau\right)} e^{j2\pi i \cdot S_{sym}Tf}, \quad (23)$$

for all $k = 0,1,\cdots,N-1$, and all $i = 0,1,\cdots,M-1$. From Eq. (23), we can conclude that the ambiguities happen at $\tau' = \tau + \frac{k_1 T_s}{S_{sub}}$, and $f' = f + \frac{k_2}{S_{sym}T} + \frac{pk_1}{S_{sub}S_{sym}T}$, where $(k_1, k_2)$ are arbitrary integers. And $(\tau, f)$ is the true (delay, Doppler frequency) pair. Fig. 3 presents EAP with (0,0) as the true (delay, Doppler) pair using 2D IAA in the case of $S_{sym} = 1, S_{sub} = 4$ with the staggering scheme A. Fig. 12(a) and 11(b) show the result of $p = 1$ and $p = 3$, respectively. The side peak locations are also the same as 2D FFT.

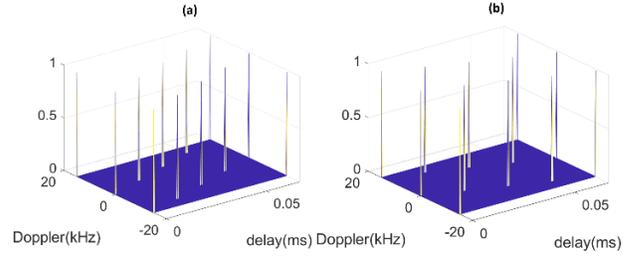

**Fig. 12.** EAP with staggering scheme D using IAA.

### 3) Proposed Staggering Scheme E for super-resolution algorithms

Note that the steering matrix $\mathbf{W}$ cannot distinguish $(\tau, f)$ and $(\tau', f') = (\tau + k_1 T_s, f + k_2/S_{sym}T)$ for any integer pair $(k_1, k_2)$, regardless of choices of staggering offsets. It is desirable to have ambiguities only at those $(\tau', f')$ values, meaning no side peaks equal to the mainlobe within the best achievable range of time delay (0 to $T_s$) and Doppler (0 to $1/S_{sym}T$). This condition for side-peak non-existence can be met if and only if the following statement is true: there are no scalars $\tau' - \tau \in (0, T_s)$ and $f' - f \in \mathbb{R}$ such that the following equation can hold (hereafter referred as "the anti-condition"):

$$e^{-j\left(\frac{2\pi(k \cdot S_{sub}+F_i)}{T_s}\tau'\right)} e^{j2\pi i \cdot S_{sym}Tf'} = e^{-j\left(\frac{2\pi(k \cdot S_{sub}+F_i)}{T_s}\tau\right)} e^{j2\pi i \cdot S_{sym}Tf}, \quad (24)$$

for all $k = 0,1,\cdots,N-1, i = 0,1,\cdots,M-1$. It can be equivalently re-written into the following form:

$$i \cdot S_{sym}T(f' - f) - (k \cdot S_{sub} + F_i)(\tau' - \tau)/T_s = \kappa_1', \quad (25)$$

where $\kappa_1'$ is an arbitrary integer for all $(i, k)$. Since the algorithm is invariant to constant phase rotation, any constant shift $\{F_i + \beta_2\}, \beta_2 \in \mathbb{Z}$ is equivalent to $\{F_i\}$. WLOG, we set $F_0 = 0$. Then, for $i = 0$, Eq. (25) gives $(\tau' - \tau) = \kappa_1 T_s/S_{sub}$. For $\tau' - \tau \in (0, T_s)$, $\kappa_1'$ must fall within $\{1,2,\cdots,S_{sub}-1\}$. Consider $i \neq 0$ and plug in $(\tau' - \tau) = \kappa_1 T_s/S_{sub}$, Eq. (25) becomes $i \cdot S_{sym}T(f' - f) - (k \cdot S_{sub} + F_i)\kappa_1/S_{sub}$. Therefore $i \cdot S_{sym}T(f' - f) - F_i\kappa_1/S_{sub}$ also needs to be an integer. Thus, the condition is equivalent to lack of a solution $(f' - f, \kappa_1)$ to the equation system

$$\mathrm{mod}(i \cdot S_{sym}S_{sub}T(f' - f) - F_i\kappa_1, S_{sub}) = 0, \quad (26)$$

for $i = 0,1,\cdots,M-1$. Note that should a solution exist, the equation for $i = 1$ implies $S_{sym}S_{sub}T(f' - f)$ is an integer, denoted as $\kappa_2$. When $F_0 = 0$, the condition is finally equivalent to no integer solutions $(\kappa_1 \in \{1,2,\cdots,S_{sub}-1\}, \kappa_2 \in \{1,\cdots,S_{sym}S_{sub}-1\})$ to the equation system

$$\mod(i\kappa_2 - F_i\kappa_1, S_{sub}) = 0, \quad (27)$$

for $i = 0,1,\cdots,M-1$. In general, for $F_0 \neq 0$, it becomes

$$\mod(i\kappa_2 - (F_i - F_0)\kappa_1, S_{sub}) = 0, \quad (28)$$

for all $i = 0,1,\cdots,M-1$. When $S_{sub}$ is prime, the only $\{F_i\}$ where Eq. (28) cannot hold is the staggering Scheme D described above. Otherwise, desirable staggering schemes based on the anti-condition can be constructed.

The design criteria for staggering Scheme E are concluded as follows: For a staggering offset sequence $\{F_i\}_{i=0}^{M-1}$, there must not exist a pair of integers $(\kappa_1, \kappa_2)$, where $(\kappa_1 \in \{1,2,\ldots,S_{sub}-1\}, \kappa_2 \in \mathbb{Z})$, that can satisfy the modulo equation system in Eq. (28). When the number of equations $M$ is less than or equal to 2, the anti-condition can always be satisfied by $\kappa_2 = (F_1 - f_0)\kappa_1$. Therefore, $M \geq 3$ is a necessary condition to guarantee non-existence of a solution to the equation system. For instance, Scheme C mentioned above satisfies the anti-condition. An example of $(M = 3, S_{sub} = 4, \{F_i\} = \{0,3,1\})$ ambiguity performance is given in Fig. 13, and another example of $S_{sub} = 4, S_{sym} = 2$ with Scheme E is given in Fig. 14, respectively. In summary, super-resolution algorithms such as IAA only need a subset of the staggering cycle to eliminate side peaks to achieve a better performance with a smaller number of RS symbols than standard resolution algorithms.

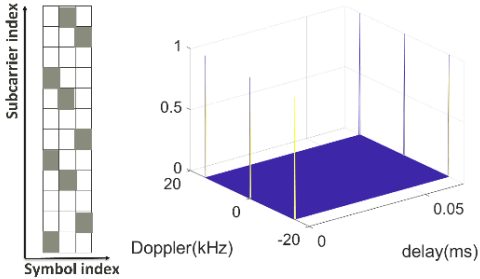

Fig. 13. EAP with staggering Scheme E ($S_{sub} = 4, S_{sym} = 1$).

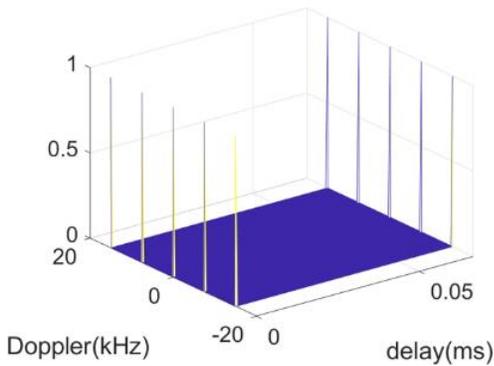

Fig. 14. EAP with staggering Scheme E ($S_{sub} = 4, S_{sym} = 2$).

## VII. SYNTHESIS OF COMB REFERENCE SIGNAL

In this section, we will investigate how to synthesize a new RS pattern for sensing by combining two existing comb RSs. The synthesized RS pattern is shown in Fig. 15, where the colors differentiate its component comb RSs. The orange pattern is the uniform symbol spacing PRS-like pattern (e.g., $S_{sub} = 4$, $S_{sym} = 3$, $\{F_0, F_1, F_2, F_3\} = \{0,2,1,3\}$ and $M = 4$). Dark blue color denotes a PTRS-like pattern. Let

1) $C_1$ (unit in subcarrier numbers) be the RE offset of the PTRS-like pattern (i.e., $C_1 = 6$ in Fig. 15),
2) $C_2$ (unit in symbol numbers) be the time-domain location of the first PTRS-like symbol (i.e., $C_2 = 1$ in Fig. 15),
3) $S_F$ (unit in subcarrier numbers) be the frequency-domain spacing of the PTRS-like REs, $S_{PT}$ (unit in symbol numbers) be the time separation of the PTRS-like symbols (i.e., $S_F = 4, S_{PT} = 3$ in Fig. 15),
4) $U_F$ be the number of frequency-domain REs for one PTRS-like symbol within one snapshot (i.e., $U_F = 2$ in Fig. 15), and
5) $U$ be the number of PTRS-like symbols within one snapshot (i.e., $U = 3$ in Fig. 15).

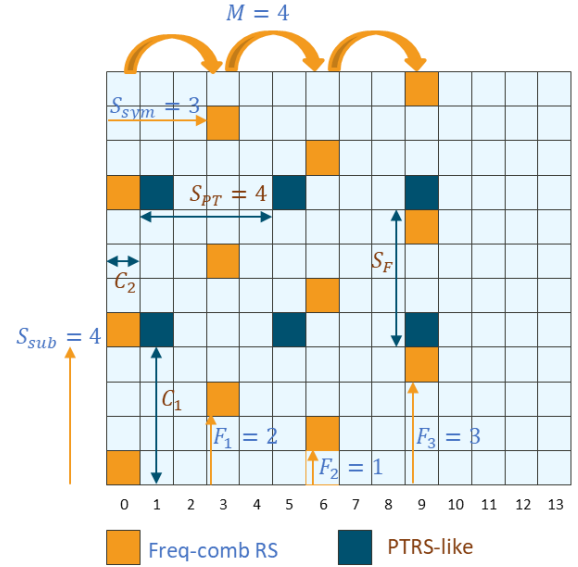

Fig. 15. Synthesized comb RS pattern.

The $h$-th column of the steering matrix (formed by MUSIC or IAA) can be written as:

$$\mathbf{w}_h(f_h, \tau_h) = \begin{bmatrix} \mathbf{w}_{h1}(f_h, \tau_h) \\ \mathbf{w}_{h2}(f_h, \tau_h) \end{bmatrix}, \quad (29)$$

where $\mathbf{w}_{h1}(f_h, \tau_h)$ is presented in Eq. (21) and

$$\mathbf{w}_{h2}(f_h, \tau_h) = \begin{bmatrix} e^{-\frac{j(2\pi C_1 \tau_h)}{T_s}} e^{j2\pi(C_2 + 0 S_{PT})T f_h} \\ e^{-\frac{j(2\pi(C_1 + S_F)\tau_h)}{T_s}} e^{j2\pi(C_2 + 0 S_{PT})T f_h} \\ \vdots \\ e^{-\frac{j(2\pi(C_1 + (U_F-1)S_F)\tau_h)}{T_s}} e^{j2\pi(C_2 + 0 S_{PT})T f_h} \\ \vdots \\ e^{-\frac{j(2\pi C_1 \tau_h)}{T_s}} e^{j2\pi(C_2 + (U-1) S_{PT})T f_h} \\ e^{-\frac{j(2\pi(C_1 + S_F)\tau_h)}{T_s}} e^{j2\pi(C_2 + (U-1) S_{PT})T f_h} \\ \vdots \\ e^{-\frac{j(2\pi(C_1 + (U_F-1)S_F)\tau_h)}{T_s}} e^{j2\pi(C_2 + (U-1) S_{PT})T f_h} \end{bmatrix}.$$

(30)

For $\mathbf{w}_{h2} = \emptyset$, analysis was given in the previous section. The best EAP is limited by $\mathbf{w}_{h1}$ not being able to distinguish $(\tau' = \tau + k_1 T_s, f' = f + k_2/(S_{sym}T))$ for integer pair $(k_1, k_2)$ from $(\tau, f)$. Now after combining with the $\mathbf{w}_{h2}$, the scheme can avoid side peaks with equal power to the main lobe between $(\tau, f)$ and $(\tau' = \tau + k_1 T_s, f' = f + k_2/T)$ for integer pair $(k_1, k_2)$. Thus, the EAP of synthesized RS can further improve that of component comb RSs, at the same time with less additional resources (reuse of communication RSs that might have already been allocated). Now we consider non-empty $w_{h2}$ (i.e., PTRS-like component included). If the vector $w_{h1}$ is empty (i.e., no PRS-like component), $w_{h2}$ alone is equivalent to non-staggered patterns, which cannot eliminate all side peaks. Therefore, we shall focus on the case of $M \geq 1$ (at least one PRS-like symbol) and $F_0 = 0$ WLOG. At the end of this section, we will apply the phase rotation back and deal with the general case. In such a situation, the pair $(\tau, f)$ cannot be distinguished from the pair $(\tau', f')$ using the information from $w_{h2}$ if the following condition holds true:

$$e^{j2\pi\left((C_2+uS_{PT})Tf - \frac{C_1+lS_F}{T_s}\tau\right)} = e^{j2\pi\left((C_2+uS_{PT})Tf' - \frac{C_1+lS_F}{T_s}\tau'\right)} \quad (31)$$

for $u = 0,1,\cdots,U-1, l = 0,1,\ldots,U_F - 1$. In the following we consider three cases.

A. *Synthesizing with PTRS and at least two staggered comb RS symbols*

In this subsection, we assume $M \geq 2$ (the PRS-like component can resolve certain Doppler shifts). Recall that in section VI. C, we have shown that, to make $(\tau, f)$ indistinguishable from the pair $(\tau', f')$, we need $\kappa_1 := (\tau' - \tau)S_{sub}/T_s$ and $\kappa_2 := S_{sym}S_{sub}T(f' - f)$ to be both integers, and the best staggering offset design of the vector $w_{h1}$ has been shown to be equivalent to Eq. (28). By substituting the expressions of $\kappa_1$ and $\kappa_2$ in Eq. (28) into Eq. (31), we get an additional anti-condition.

$$\mathrm{mod}\big((C_2+uS_{PT})\kappa_2 - (C_1+lS_F)S_{sym}\kappa_1, S_{sym}S_{sub}\big)=0, \quad (32)$$

for any $u = 0,1,\cdots,U-1, l=0,1,\ldots,U_F-1$. Above derivations assume $F_0 = 0$. To generalize, one may replace $C_1$ with $C_1 - F_0$, and $F_i$ with $F_i - F_0$ in Eq. (32). And the general rules for combining staggering frequency comb RS with a PTRS-like pattern is: neither the condition in Eq. (28) nor the following anti-condition of PTRS-like pattern

$$\mathrm{mod}\big((C_2+uS_{PT})\kappa_2 - (C_1+lS_F-F_0)S_{sym}\kappa_1, S_{sym}S_{sub}\big)=0, \quad (33)$$

holds true for $u = 0,1,\cdots,U-1, l = 0,1,\cdots,U_F - 1$.

Then the ambiguity due to aliasing does not exist in desired ranges of delay $\tau \in (0, T_s)$ and Doppler $f \in (0, 1/T)$. In doing that, we need to guarantee that no non-zero integer solution $(\kappa_1, \kappa_2) \in \{0,1,2,\cdots,S_{sub}-1\} \times \{0,1,\cdots,S_{sym}S_{sub}-1\}$ to the two anti-conditions Eq. (28) and (33) exists. The anti-condition Eq. (33) can be simplified under different situations.

- If $U \geq 2$ and $U_F \geq 2$, Eq. (33) is equivalent to the following set of conditions:
  $\mathrm{mod}(S_{PT}\kappa_2, S_{sym}S_{sub}) = 0$, $\mathrm{mod}(S_F\kappa_1, S_{sub}) = 0$, and $\mathrm{mod}(C_2\kappa_2 - (C_1 - F_0)\kappa_1, S_{sym}S_{sub}) = 0$,
- If $U \geq 2$ and $U_F = 1$, Eq. (33) is equivalent to the following set of conditions: $\mathrm{mod}(S_{PT}\kappa_2, S_{sym}S_{sub}) = 0$, $\mathrm{mod}(C_2\kappa_2 - (C_1 - F_0)\kappa_1, S_{sym}S_{sub}) = 0$.
- If $U = 1$ and $U_F \geq 2$, Eq. (33) is equivalent to the following set of conditions: $\mathrm{mod}(S_F\kappa_1, S_{sub}) = 0$, $\mathrm{mod}(C_2\kappa_2 - (C_1 - F_0)\kappa_1, S_{sym}S_{sub}) = 0$.

Here are some examples. Consider the case of $M = 2$, with $F_0 = 0$ and $F_1 = 1$. One way of ensuring the nonexistence of the solution is by taking $U \geq 1$, and choosing the pair $(C_1, C_2)$ such that $C_2 + u^* S_{PT} - (C_1 + l^* S_F - F_0)S_{sym}$ and $S_{sub}$ are relatively prime for some $u^* \in \{0,1,\cdots,U-1\}$ and $l^* \in \{0,1,\cdots,U_F - 1\}$. In such case, first anti-condition Eq. (28) implies that $\kappa_2 - \kappa_1$ is an integer multiple of $S_{sub}$, which implies that

$$\mathrm{mod}\big((C_1 + lS_F - F_0)S_{sym}(\kappa_1 - \kappa_2), S_{sym}S_{sub}\big) = 0. \quad (34)$$

Substituting the above equation into the anti-condition Eq. (33), we have

$$\mathrm{mod}\big((C_2 + uS_{PT} - (C_1 + lS_F - F_0)S_{sym})\kappa_2, S_{sym}S_{sub}\big) = 0. \quad (35)$$

Since we have $C_2 + u^* S_{PT} - (C_1 + l^* S_F - F_0)S_{sym}$ and $S_{sym}S_{sub}$ are relatively prime for some $u^* \in \{0,1,\cdots,U-1\}$ and $l^* \in \{0,1,\cdots,U_F - 1\}$, the anti-condition at the pair $(l^*, u^*)$ requires $\kappa_2$ to be an integer multiple of $S_{sym}S_{sub}$, which implies that $\kappa_1$ is an integer multiple of $S_{sub}$. Since we have $\kappa_1 \in \{0,1,\cdots,S_{sub}-1\}$ and $\kappa_2 \in \{0,1,\cdots,S_{sym}S_{sub} - 1\}$, the only possibility is $\kappa_1 = \kappa_2 = 0$. Therefore, under this setup, when we have $U \geq 1$ and $C_2 + u^* S_{PT} - (C_1 + l^* S_F - F_0)S_{sym}$ and $S_{sym}S_{sub}$ are relatively prime for some $u^* \in \{0,1,\cdots,U-1\}$ and $l^* \in \{0,1,\cdots,U_F - 1\}$, the ambiguity due to aliasing does not exist in the desired range.

Fig. 16(a) shows the results of $M = 2, N = 15, F_0 = 0$ and $F_1 = 1$, $S_{sub} = 6$, $C_1 = 1$, $C_2 = 3$, $S_{sym} = 4$, $U = 1$, $U_F = 1$ using IAA. Fig. 16(b) illustrates $M = 2$, $F_0 = 0$ and $F_1 = 1$, $S_{sub} = 4$, $C_1 = 1$, $C_2 = 3$, $S_{sym} = 10$, $U = 1$, $U_F = 3$, $S_F = 3$ using IAA. Compared with Fig. 15, the EAP in Fig.15 is improved by incorporating irregular RS pattern, even with smaller number of REs.

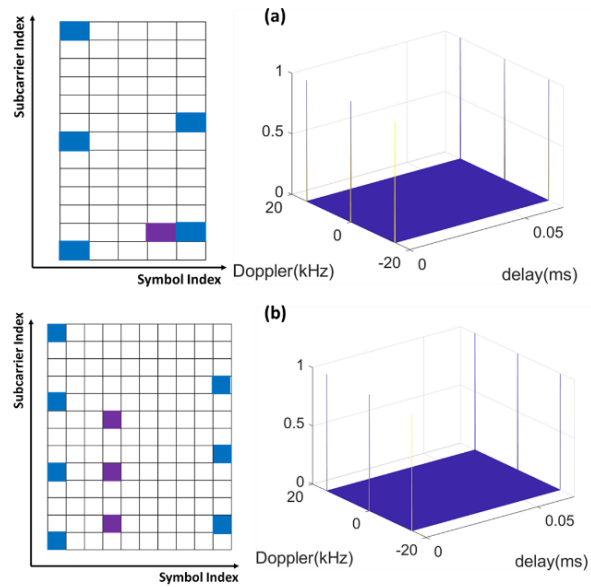

**Fig. 16**. Examples of synthesizing PTRS with at least two Staggered Comb RS Symbols.

## B. Synthesizing with at least two PTRS symbols and one comb RS symbol

This case assumes $M = 1$, $U \geq 2$. First, we consider a case where there is only one RE per PTRS symbol (i.e., $U_F = 1$). The necessary and sufficient condition is the non-existence of a non-zero integer solution $(\kappa_1, \kappa_3) \in [0, S_{sub} - 1] \times [0, S_{PT} - 1]$ to the following anti-condition:

$$\frac{C_2}{S_{PT}}\kappa_3 - \frac{C_1 - F_0}{S_{sub}}\kappa_1 \in \mathbb{Z} \quad (36)$$

must be satisfied. Derivation of Eq. (36) is shown in Appendix, Section A.

Let us discuss some of these conditions in more detail. First, note that the pair $(\tau' = \tau + \frac{T_s}{\gcd(C_1, S_{sub})}, f' = f)$ and $(\tau, f)$ make the anti-condition Eq. (36) hold true, where gcd () is the greatest common divisor. To ensure that no solution exists for $\tau' - \tau \in (0, T_s)$, $C_1$ and $S_{sub}$ shall be relatively prime. On the same token, the pair $(\tau' = \tau, f' = f + \frac{1}{\gcd(C_2, S_{PT})T})$ and $(\tau, f)$ make the anti-condition hold true. To ensure that no solution exists for $f' - f \in (0, 1/T)$, $C_2$ and $S_{PT}$ shall be relatively prime. These two conditions are necessary but not sufficient. On the other hand, one can construct examples for which the anti-condition still holds. Suppose we design the system such that $\gcd(S_{PT}, S_{sub}) = 1$, then the quantity $\frac{C_2}{S_{PT}}\kappa_3 - \frac{C_1}{S_{sub}}\kappa_1$ being an integer implies that both terms are integers. As long as we ensure that $C_1$ and $S_{sub}$ are relatively prime, and that $C_2$ and $S_{PT}$ are relative prime, the non-existence of the solution to Eq. (36) can be ensured. Fig. 17 illustrates an example of a synthesized RS scheme with the right combinations of comb-RS and PTRS parameters ($M = 1$, $N = 15$, $F_0 = 0$, $S_{sub} = 2, C_1 = 1, C_2 = 1$, $U = 3$, $U_F = 1, S_{PT} = 1$) using IAA.

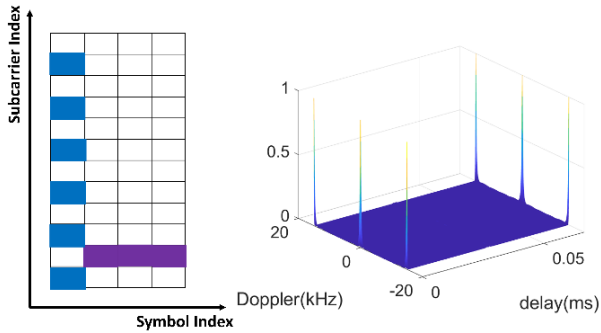

**Fig. 17.** An example of synthesizing at least two PTRS symbols ($U_F = 1$) with one comb RS symbol.

We now investigate the case of $U_F \geq 2$ (multi-tone PTRS-like component in Fig. 18). In order to ensure the non-existence of side peaks in the range $\tau' - \tau \in (0, T_s), f' - f \in \left(0, \frac{1}{T}\right)$, the necessary and sufficient condition is: non-zero integer solution $(\kappa_1, \kappa_3) \in [0, S_{sub} - 1] \times [0, S_{PT} - 1]$ to the following anti-conditions

$$\frac{C_2}{S_{PT}}\kappa_3 - \frac{C_1 - F_0}{S_{sub}}\kappa_1 \in \mathbb{Z}, \quad \frac{S_F}{S_{sub}}\kappa_1 \in \mathbb{Z}. \quad (37)$$

does not exist. Section B of Appendix illustrates the detailed derivation of Eq. (37). Fig. 18 illustrates an example where $M = 1$, $N = 15$, $F_0 = 0$, $S_{sub} = 5, C_1 = 1, C_2 = 1$, $U = 3$, $U_F = 2, S_{PT} = 3, S_F = 3$ using IAA.

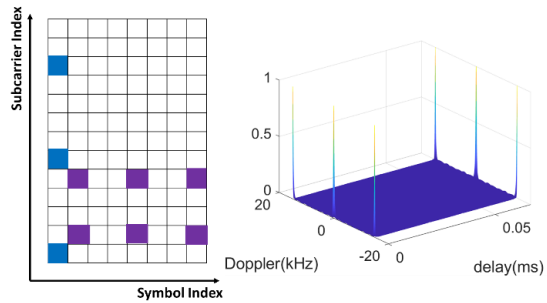

**Fig. 18.** An example of synthesizing at least two PTRS symbols ($U_F = 2$) with one comb RS symbol.

## C. Synthesizing with one PTRS symbol and one comb RS symbol

In this case, $M = 1$, $U = 1$ and we need $U_F \geq 2$ to eliminate ambiguities. In order to ensure the non-existence of side peaks in the range $\tau' - \tau \in (0, T_s), f' - f \in \left(0, \frac{1}{T}\right)$, the necessary and sufficient condition is: non-zero integer solution $(\kappa_1, \kappa_3) \in [0, S_{sub} - 1] \times [0, S_{PT} - 1]$ to the following anti-conditions

$$\mod(\kappa_4 - (C_1 - F_0)\kappa_1, S_{sub}) = 0, \quad \frac{S_F}{S_{sub}}\kappa_1 \in \mathbb{Z}. \quad (38)$$

does not exist. Section C of Appendix shows the derivation of Eq. (38). Fig. 19 illustrates an example where $M = 1$, $N = 15, F_0 = 0, S_{sub} = 5, C_1 = 1, C_2 = 1, U = 1, U_F = 4, S_F = 3$.

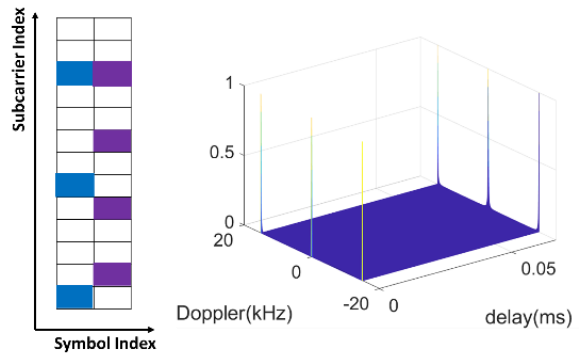

**Fig. 19.** An example of synthesizing one PTRS symbol with one comb RS symbol.

## VIII. GENERAL IRREGULAR RS PATTERNS

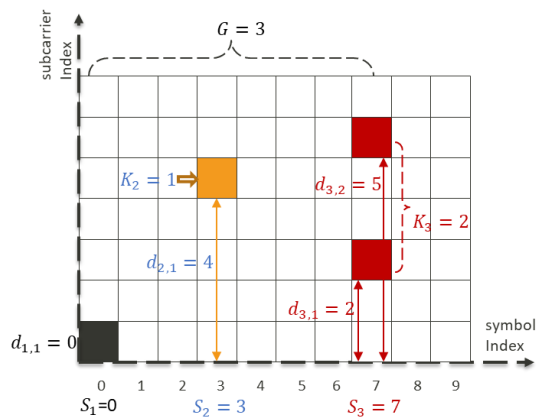

**Fig. 20.** General irregular RS pattern.

Now let's turn to general irregular RS pattern. Denote
1) $d_{g,j}$ as RE offset of the $j$th RE at the $g$th RS symbol,
2) $K_g$ as the total number of RS REs at the $g$th RS symbol,
3) $S_g$ as the RS symbol index of the $g$th RS symbol, and
4) $G$ as the total number of RS symbols.

A graphical illustration is given in Fig. 20. Note that $d_{g,j}$ and $S_g$ are not necessarily uniform spaced, which could form an irregular pattern. The $h$-th column of the steering matrix is therefore listed below:

$$\mathbf{w}'_h(f_h, \tau_h) = \begin{bmatrix} e^{-\frac{j(2\pi d_{11}\tau_h)}{T_s}} e^{j2\pi S_1 T f_h} \\ e^{-\frac{j2\pi d_{12}\tau_h}{T_s}} e^{j2\pi S_1 T f_h} \\ \vdots \\ e^{-\frac{j2\pi d_{1K_1}\tau_h}{T_s}} e^{j2\pi S_1 T f_h} \\ e^{-\frac{j(2\pi d_{21}\tau_h)}{T_s}} e^{j2\pi S_2 T f_h} \\ e^{-\frac{j2\pi d_{22}\tau_h}{T_s}} e^{j2\pi S_2 T f_h} \\ \vdots \\ e^{-\frac{j2\pi d_{2K_2}\tau_h}{T_s}} e^{j2\pi S_2 T f_h} \\ \vdots \\ e^{-\frac{j(2\pi d_{G1}\tau_h)}{T_s}} e^{j2\pi S_G T f_h} \\ e^{-\frac{j(2\pi d_{G2}\tau_h)}{T_s}} e^{j2\pi S_G T f_h} \\ \vdots \\ e^{-\frac{j(2\pi d_{GK_G}\tau_h)}{T_s}} e^{j2\pi S_G T f_h} \end{bmatrix}, \quad (39)$$

A pair of parameters $(\tau, f)$ results in ambiguity with another pair $(\tau', f')$, if for any $g \in \{1,2,\cdots,G\}$ and $k \in \{1,2,\cdots,K_g\}$,

$$\Phi(g,k) = (S_g - S_1)T(f' - f) - (d_{g,k} - d_{1,1})\frac{\tau'-\tau}{T_s} \in \mathbb{Z}. \quad (40)$$

Conditions to avoid ambiguities are given next.

*A. Two or more REs for at least one RS symbol*

In this case, there exists at least one $g^* \in \{1,2,\cdots,G\}$, such that $K_{g^*} \geq 2$. Take an arbitrary distinct integer pair $(k^*, l^*) \in \{1,\cdots,K_{g^*}\} \times \{1,\cdots,K_{g^*}\}$, where $k^* \neq l^*$. And for the sake of computational simplicity, one may choose $d^* \coloneqq \min_{g:K_g \geq 2} \min_{k \neq l} |d_{g,k} - d_{g,l}|$.) Due to invariance to constant phase rotation, we can assume $d_{11} = S_1 = 0$ WLOG. Applying Eq. (40) with $(g^*, k^*)$ and $(g^*, l^*)$ and taking their difference, we define $\kappa_1 \coloneqq d^*\frac{\tau'-\tau}{T_s} \in \mathbb{Z}$. For an ambiguity to exist in the range $\tau' - \tau \in [0, T_s]$, we have $\kappa_1 \in \{0,1,2,\cdots,d^*-1\}$. Substituting into Eq (40), we have

$$S_g T(f' - f) - \frac{d_{g,k}}{d^*}\kappa_1 \in \mathbb{Z}, \quad (41)$$

for any $g \in \{1,2,\cdots,G\}$ and $k \in \{1,2,\cdots,K_g\}$. Consequently, $d^* S_g T(f' - f) \in \mathbb{Z}$. Since we assume $S_1 = 0$, we need $G \geq 2$ and $S_2 \neq 0$ for this constraint not to be trivial. Let $S^* \coloneqq \gcd(S_2 - S_1, \cdots, S_G - S_1)$, we have $\kappa_2 \coloneqq S^* d^* T(f' - f) \in \mathbb{Z}$. If $f' - f \in [0, 1/T]$, we have $\kappa_2 \in \{0,1,2,\cdots,S^*d^*-1\}$. Under such a representation, the anti-condition Eq. (40) becomes: $\mathrm{mod}(S_g\kappa_2 - d_{g,k}S^*\kappa_1, S^*d^*) = 0$, for any $g \in \{1,2,\cdots,G\}$ and $k \in \{1,2,\cdots,K_g\}$.

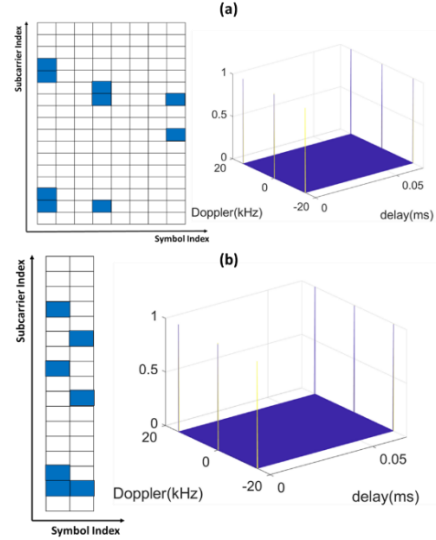

**Fig. 21**. Examples of irregular RS design based on two or more REs for at least One RS symbol.

Above derivations assumed $S_1 = d_{1,1} = 0$. In general, one may replace $S_g$ with $S_g - S_1$, and $d_{g,k}$ with $d_{g,k} - d_{1,1}$, and obtain the anti-condition

$$mod\left((S_g - S_1)\kappa_2 - (d_{g,k} - d_{1,1})S^*\kappa_1, S^*d^*\right) = 0, \quad (42)$$

for any $g \in \{1,2,\cdots,G\}$ and $k \in \{1,2,\cdots,K_g\}$.

Therefore, the distinguishability in the desirable range $\tau' - \tau \in (0, T_s)$, $f' - f \in (0, 1/T)$ is equivalent to: there exist no $(\kappa_1, \kappa_2)$ solutions to Eq (42) other than $(0,0)$, where $\kappa_1 \in \{0,1,\cdots,d^*-1\}$, $\kappa_2 \in \{0,1,\cdots,S^*d^*-1\}$. In this case, we need $G \geq 2$ and at least one RS symbol satisfying $K_g \geq 2$ such that least 3 entries exist in the vector $\mathbf{w}'_h$. Fig. 21(a) shows the result of $G = 3$, $d_{1,1} = 1$, $d_{1,2} = 2$, $d_{1,3} = 12$, $d_{1,4} = 13$, $d_{2,1} = 1$, $d_{2,2} = 10$, $d_{2,3} = 11$, $d_{3,1} = 7$, $d_{3,2} = 10$, $S_1 = 0$, $S_2 = 3$, $S_3 = 7$. Fig. 21(b) presents the result of $G = 2$, $d_{1,1} = 1$, $d_{1,2} = 2$, $d_{1,3} = 9$, $d_{1,4} = 13$, $d_{2,1} = 1$, $d_{2,2} = 7$, $d_{2,3} = 11$, $S_1 = 0$, $S_2 = 1$.

*B. Single RE for every RS symbol*

This RS pattern is parameterized as: $\forall g \in \{1,2,\cdots,G\}$, $K_g = 1$ and it is analogous to a frequency hopping signal. For notation simplicity, we use $d_g$ to denote $d_{g,1}$ for any $g \in \{1,2,\cdots,G\}$. In such case, we need the existence of a pair of integers $m^*, n^* \geq 2$, $m^* \neq n^*$, such that $\frac{d_{m^*}-d_1}{S_{m^*}-S_1} \neq \frac{d_{n^*}-d_1}{S_{n^*}-S_1}$. Suppose that such a pair does not exist, it is equivalent to stating that all $(d_g - d_1, S_g - S_1)$, $g \in \{1,2,\cdots,G\}$, are integer multiples of an integer pair $(\tilde{d}, \tilde{S})$. Any $(\tau, f)$ pair satisfying $\frac{\tau'-\tau}{T_s}(\tilde{d} - d_1) = (\tilde{S} - S_1)T(f' - f)$ will lead to ambiguity. Therefore, we need the existence of such $(m^*, n^*)$. Moreover, applying Eq (40) with $(m^*, 1)$ and $(n^*, 1)$ separately and combining the results together, we have

$$(S_{n^*} - S_1) \times \Phi(m^*, 1) - (S_{m^*} - S_1) \times \Phi(n^*, 1) \in \mathbb{Z}. \quad (43)$$

TABLE I
Summary of RS pattern design criteria for maximum achievable unambiguous range

| Sensing algorithms | RS Pattern | Proposed design criterion |
|---|---|---|
| Delay and sum | Comb RS | $F_i = \mod(p \cdot (i-1), S_{sub})$ for $i = 0, \cdots, S_{sub} - 1$, where $p$ is relatively prime to $S_{sub}$. |
| 2D FFT | Comb RS | $F_i = \mod(p \cdot (i-1), S_{sub})$ for $i = 0, \cdots, S_{sub} - 1$, where $p$ is relatively prime to $S_{sub}$. |
| Super-resolution algorithms | Comb RS | No integer solutions ($\kappa_1 \in \{1,2,\cdots, S_{sub} - 1\}, \kappa_2 \in \mathbb{Z}$) to $\mod(i \kappa_2 - (F_i - F_0)\kappa_1, S_{sub}) = 0$. |
| Super-resolution algorithms | Synthesizing PTRS with at least two staggered comb RS symbols | Non-zero integer solution $(\kappa_1, \kappa_2) \in \{0,1,2,\cdots, S_{sub} - 1\} \times \{0,1,\cdots, S_{sym}S_{sub} - 1\}$ to the two anti-conditions: Eq. (28) and (33) |
| Super-resolution algorithms | Synthesizing at least two PTRS symbols with one comb RS symbol | non-existence of a non-zero integer solution $(\kappa_1, \kappa_3) \in [0, S_{sub} - 1] \times [0, S_{PT} - 1]$ to Eq. (36) |
| Super-resolution algorithms | Synthesizing one PTRS Symbol with one comb RS symbol | non-zero integer solution $(\kappa_1, \kappa_3) \in [0, S_{sub} - 1] \times [0, S_{PT} - 1]$ to Eq. (38) |
| Super-resolution algorithms | Irregular RS: Two or more REs for at least one RS symbol | A non-zero $(\kappa_1, \kappa_2)$ to Eq (42), where $\kappa_1 \in \{0,1,\cdots, d^* - 1\}, \kappa_2 \in \{0,1,\cdots, S^*d^* - 1\}$, does not exist. |
| Super-resolution algorithms | Irregular RS: Single RE for every RS symbol | No non-zero integer pair $(\kappa_1, \kappa_2)$, where $\kappa_1, \kappa_2 \in \{0,1,\cdots, S^*_{m^*,n^*} - 1\})$ to satisfy Eq. (44). |

Define $S^*_{m^*,n^*} = |(d_{m^*} - d_1)(S_{n^*} - S_1) - (S_{m^*} - S_1)(d_{n^*} - d_1)|$, one obtains $\kappa_1 \coloneqq S^*_{m^*,n^*} \frac{\tau' - \tau}{T_s} \in \mathbb{Z}$. On the other hand, $(d_{n^*} - d_1) \times \Phi(m^*, 1) - (d_{m^*} - d_1) \times \Phi(n^*, 1) \in \mathbb{Z}$ yileds $\kappa_2 \coloneqq S^*_{m^*,n^*} T(f' - f) \in \mathbb{Z}$. Substituting $\kappa_1$ and $\kappa_2$ into Eq. (40), the anti-condition is: there is no non-zero integer pair $(\kappa_1, \kappa_2)$, where $\kappa_1, \kappa_2 \in \{0,1,\cdots, S^*_{m^*,n^*} - 1\})$, such that, for any $g$, $m^*$, $n^* \in \{1,2,\cdots, G\}$

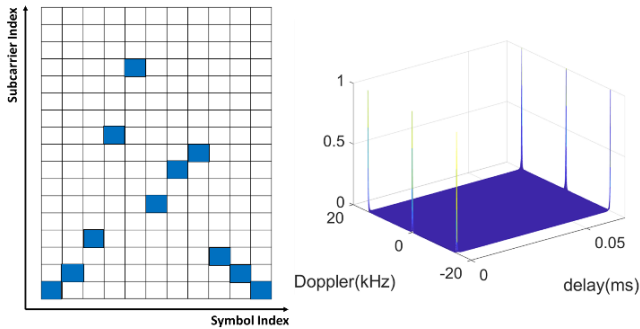

**Fig. 22.** An example of irregular RS design based on single RE for every RS symbol.

$$\mod\left((S_g - S_1)\kappa_2 - (d_g - d_1)\kappa_1, S^*_{m^*,n^*}\right) = 0. \quad (44)$$

Therefore, we have $G \geq 3$, and at least 3 entries are there in the vector $\mathbf{w}'_h$. Fig. 22 shows an instance with $G = 11$, $d_1 = 0$, $d_2 = 1$, $d_3 = 3$, $d_4 = 9$, $d_5 = 13$, $d_6 = 5$, $d_7 = 7$, $d_8 = 8$, $d_9 = 2$, $d_{10} = 1$, $d_{11} = 0$, and $S_g = g - 1$ for $g = 1, 2, 3, \cdots, 11$.

### IX. PERFORMANCE EVALUATION

Proposed design criteria in various cases are summarized in Table I. The percentage of sensing RE overheads of previous design examples achieving the EAP is listed in Table II. The results indicate that the irregular RS pattern could significantly reduce the sensing overhead while attaining the best achievable EAP.

TABLE II
BENCHMARKS OF SENSING OVERHEAD

| Sensing RS pattern | Sensing RS overhead |
|---|---|
| Full OFDM signal | 100% |
| RS pattern in Fig. 13 | 25% |
| RS Pattern in Fig. 16(a) | 6.7% |
| RS pattern in Fig. 17 | 13% |
| RS pattern in Fig. 18 | 3% |
| RS pattern in Fig. 21(a) | 8% |
| RS pattern in Fig. 22 | 7.7% |

Simulation results of multiple-target detection with a comb-6 RS example are presented below. Since 2D FFT and Delay-and-Sum show similar EAP, only results of 2D FFT are given in this section. Four targets with different locations and velocities are shown in red-circles in Fig. 23. In Fig. 23 (a) and (b), one can observe that, for standard-resolution algorithm, the proposed design can eliminate most side peaks in time delay along the zero Doppler frequency compared with non-staggered RS (RS pattern in Fig. 3(a)). Also, 2D FFT using the conventional CP fails to detect the fourth target due to interference as shown in Fig. 23 (c). As for super-resolution algorithms shown in Fig. 23(d) and (e) (with IAA as an example), the proposed design is sufficient to eliminate all side peaks and achieves the same EAP as the full OFDM signal. Moreover, Fig. 23(f) shows that super-resolution algorithms are more sensitive to ISI and exhibit worse performance degradation without the extended guard interval design.

### X. CONCLUSION

In this paper, we derived general rules for RS design with both standard- and super-resolution sensing algorithms in a communication-centric, bi-static ISAC system. The design criterion is to reduce the ambiguities (i.e., enlarge the 2D maximum unambiguity), or to make side peak locations in delay and Doppler frequency domains as far away from the mainlobe as possible. First, we investigated uniformly spaced comb RS patterns with various staggering offsets for standard-resolution algorithms. In that case, a staggering scheme with linear slope relatively prime to the comb size offers more flexible choices of 2D unambiguity. Second, we proposed an extended guard interval design for comb RS structure when us-ing post-FFT sensing algorithms to support long-range sensing with ISI beyond the OFDM CP. Third, we proposed a general comb RS staggering offset design specified for super-resolution sensing algorithms, improving ambiguity performance over standard-resolution algorithm with yet smaller RS resource usage. Furthermore, we synthesized comb RS patterns and generalized to irregular RSs for better efficiency when using super-resolution sensing algorithms. By doing so, we achieved the optimal ambiguity performance as if all time and frequency resources of OFDM signals were allocated. The results can be applied to extend the sensing distance and velocity range with OFDM RS patterns either stemming from existing communication system or starting anew in the next generation.

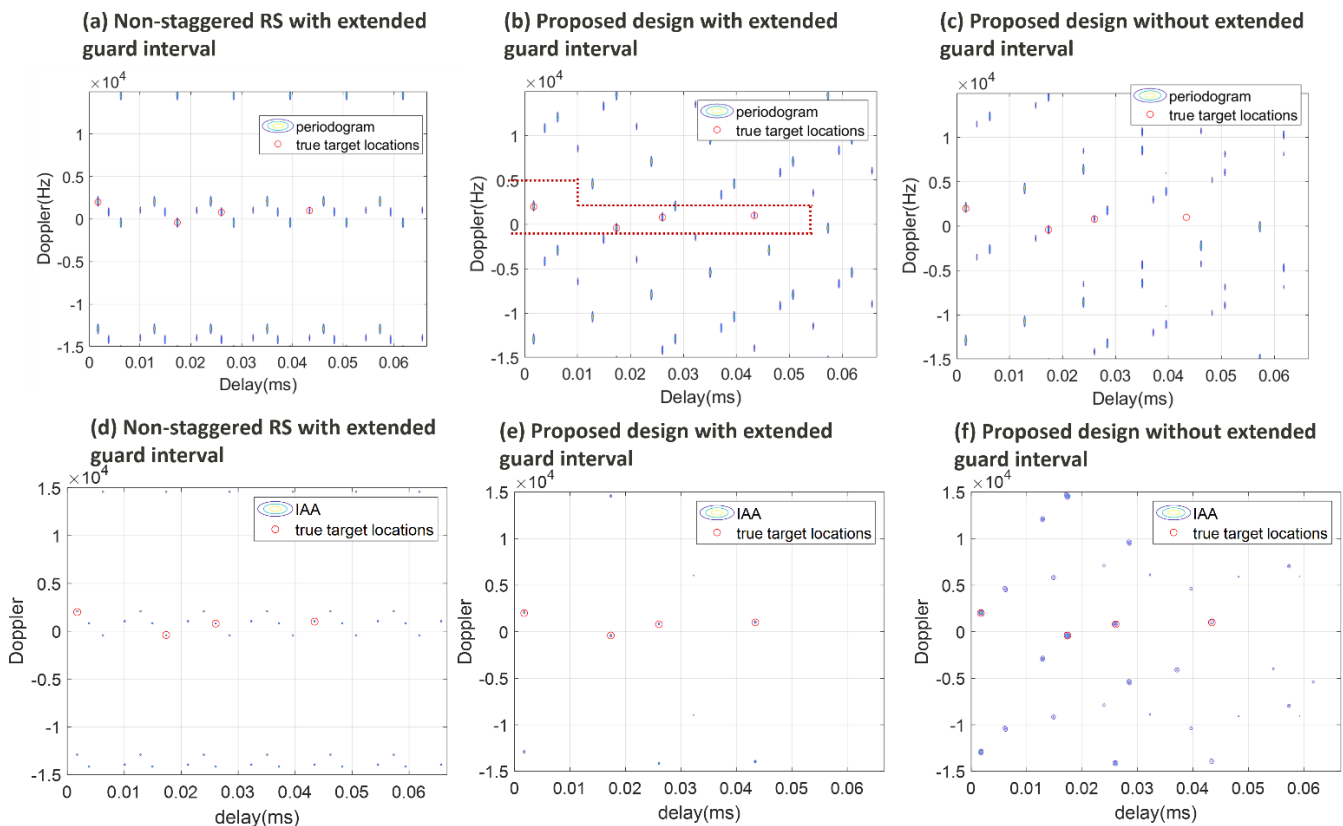

**Fig. 23.** Benchmarks of different designs for multiple targets detection. (a) to (c) are 2D FFT and (d) to (f) are super-resolution algorithms IAA.

## APPENDIX

### A. Derivation of Eq. (36)

The derivation is based on $M = 1, U \geq 2, U_F = 1$. Comparing $(u = 0, l = 0)$ and $(u = 1, l = 0)$ for Eq (31) implies that ambiguity arises in $\mathbf{w}_{h2}$ if $\kappa_3 = S_{PT}T(f' - f)$ is an integer in $\{0,1,\cdots,S_{PT} - 1\}$. Additionally, the indistinguishability of the vector $\mathbf{w}_{h1}$ in the $u = 0$ case requires $\kappa_1 := (\tau' - \tau)S_{sub}/T_s$ to be an integer in $\{0,1,\cdots,S_{sub} - 1\}$. Furthermore, applying Eq (31) with $(u = 0, l = 0)$ implies ambiguity if the term $C_2 T(f' - f) - \frac{C_1}{T_s}(\tau' - \tau) = \frac{C_2}{S_{PT}}\kappa_3 - \frac{C_1}{S_{sub}}\kappa_1$ is an integer. On the other hand, if $\kappa_1, \kappa_3, \frac{C_2}{S_{PT}}\kappa_3 - \frac{C_1}{S_{sub}}\kappa_1$ are all integers, then $(C_2 + u\,S_{PT})T(f' - f) - \frac{C_1}{T_s}(\tau' - \tau) = \frac{C_2}{S_{PT}}\kappa_3 - \frac{C_1}{S_{sub}}\kappa_1 + l\kappa_3$ is an integer and the anti-condition Eq. (31) will hold for any $u \geq 0$. Therefore, for a single comb RS symbol's ambiguities to be resolved by a PTRS-tone ($U \geq 2, U_F = 1$), the necessary and sufficient condition is: non-zero integer solution $(\kappa_1, \kappa_3) \in [0, S_{sub} - 1] \times [0, S_{PT} - 1]$ to the following anti-condition:

$$\frac{C_2}{S_{PT}}\kappa_3 - \frac{C_1}{S_{sub}}\kappa_1 \in \mathbb{Z}. \quad (45)$$

does not exist. Above derivations assume $F_0 = 0$. In general, one may replace $C_1$ with $C_1 - F_0$, and obtain the anti-condition, i.e.,

$$\frac{C_2}{S_{PT}}\kappa_3 - \frac{C_1 - F_0}{S_{sub}}\kappa_1 \in \mathbb{Z} \quad (46)$$

must be satisfied.

### B. Derivation of Eq. (37)

The derivation is based on $M = 1, U \geq 2, U_F > 1$. Similar to derivation with $U_F = 1$, the existence of side peaks implies that $\kappa_3 = S_{PT}T(f' - f)$ is an integer in $\{0,1,\cdots,S_{PT} - 1\}$, $\kappa_1 := (\tau' - \tau)\frac{S_{sub}}{T_s}$ is an integer in $\{0,1,\cdots,S_{sub} - 1\}$, and that $\frac{C_2}{S_{PT}}\kappa_3 - \frac{C_1}{S_{sub}}\kappa_1$ is an integer. Furthermore, comparing the cases $(l = 1, u = 0)$ and $(l = 0, u = 0)$ for Eq (31), we obtain that $\frac{S_F}{T_s}(\tau' - \tau) = \frac{S_F}{S_{sub}}\kappa_1$ is an integer. On the other hand, if $\kappa_1, \kappa_3, \frac{C_2}{S_{PT}}\kappa_3 - \frac{C_1}{S_{sub}}\kappa_1, \frac{S_F}{S_{sub}}\kappa_1$ are all integers, we have

$$(C_2 + l\,S_{PT})T(f' - f) - \frac{C_1 + uS_F}{T_s}(\tau' - \tau)$$

$$= \frac{C_2}{S_{PT}}\kappa_3 - \frac{C_1}{S_{sub}}\kappa_1 + l\kappa_3 - \frac{S_F}{S_{sub}}\kappa_3 \in \mathbb{Z}. \quad (47)$$

and the anti-condition Eq (31) will hold for any $l \geq 0$ and $i \geq 0$. Therefore, in order to ensure the non-existence of side peaks in the range $\tau' - \tau \in (0, T_s), f' - f \in \left(0, \frac{1}{T}\right)$, the necessary and sufficient condition is: non-zero integer solution $(\kappa_1, \kappa_3) \in [0, S_{sub} - 1] \times [0, S_{PT} - 1]$ to the following anti-conditions:

$$\frac{C_2}{S_{PT}}\kappa_3 - \frac{C_1}{S_{sub}}\kappa_1 \in \mathbb{Z}, \quad \frac{S_F}{S_{sub}}\kappa_1 \in \mathbb{Z}. \quad (48)$$

does not exist. Above derivations assume $F_0 = 0$. In general, one may replace $C_1$ with $C_1 - F_0$, and $F_i$ with $F_i - F_0$, and obtain the anti-condition:

$$\frac{C_2}{S_{PT}}\kappa_3 - \frac{C_1 - F_0}{S_{sub}}\kappa_1 \in \mathbb{Z}, \quad \frac{S_F}{S_{sub}}\kappa_1 \in \mathbb{Z}. \quad (49)$$

### C. Derivation of Eq. (38)

In this case, we assume $M = 1, U = 1$ and we need $U_F \geq 2$ to eliminate ambiguities. The indistinguishability of the vector $\boldsymbol{w}_{h1}$ in the $i = 0$ case implies that $\kappa_1 := (\tau' - \tau)\frac{S_{sub}}{T_s}$ is an integer in $\{0, 1, \cdots, S_{sub} - 1\}$. Comparing the cases $(l = 1, u = 0)$ and $(u = 0, l = 0)$ for Eq (31), we obtain that $\frac{S_F}{T_s}(\tau' - \tau) = \frac{S_F}{S_{sub}}\kappa_1$ is an integer. Furthermore, applying Eq (31) with $(u = 0, l = 0)$ implies that $C_2 T(f' - f) - \frac{C_1}{T_s}(\tau' - \tau) = C_2 T(f' - f) - \frac{C_1}{S_{sub}}\kappa_1$ is an integer. So we have $\kappa_4 := C_2 S_{sub} T(f' - f)$ is an integer. In such case, the necessary and sufficient condition for the non-existence of ambiguous side peaks is: non-zero integer solution $(\kappa_1, \kappa_4) \in [0, S_{sub} - 1] \times [0, C_2 S_{sub} - 1]$ to the following anti-conditions:

$$\mathrm{mod}(\kappa_4 - C_1\kappa_1, S_{sub}) = 0, \quad \frac{S_F}{S_{sub}}\kappa_1 \in \mathbb{Z}. \quad (50)$$

does not exist. Note that if $C_2 > 1$, such an anti-condition is always satisfied by $(\kappa_1 = 0, \kappa_4 = S_{sub})$. Therefore, for the non-existence of a solution to hold, we need $C_2 = 1$. Taking into account the constant phase rotation caused by $F_0$, the necessary and sufficient condition is the lack of a non-zero integer solution $(\kappa_1, \kappa_3) \in [0, S_{sub} - 1] \times [0, S_{PT} - 1]$ to the following anti-conditions:

$$\mathrm{mod}(\kappa_4 - (C_1 - F_0)\kappa_1, S_{sub}) = 0, \quad \frac{S_F}{S_{sub}}\kappa_1 \in \mathbb{Z}. \quad (51)$$

### REFERENCES

[1] Z. Wei, H. Qu, Y. Wang, X. Yuan, H. Wu, Y. Du, K. Han, N. Zhang, and Z. Feng, "Integrated Sensing and Communication Signals Toward 5G-A and 6G: A Survey," *IEEE Internet of Things Journal,* vol. 10, no. 13, pp. 11068-11092, 2023.
[2] S. Lu, F. Liu, Y. Li, K. Zhang, H. Huang, J. Zou, X. Li, Y. Dong, F. Dong, J. Zhu, Y. Xiong, W. Yuan, Y. Cui, and L. Hanzo, "Integrated Sensing and Communications: Recent Advances and Ten Open Challenges," *IEEE Internet of Things Journal,* pp. 1-1, 2024.
[3] F. Liu, Y. Cui, C. Masouros, J. Xu, T. X. Han, Y. C. Eldar, and S. Buzzi, "Integrated Sensing and Communications: Toward Dual-Functional Wireless Networks for 6G and Beyond," *IEEE Journal on Selected Areas in Communications,* vol. 40, no. 6, pp. 1728-1767, 2022.
[4] J. Wang, N. Varshney, C. Gentile, S. Blandino, J. Chuang, and N. Golmie, "Integrated Sensing and Communication: Enabling Techniques, Applications, Tools and Data Sets, Standardization, and Future Directions," *IEEE Internet of Things Journal,* vol. 9, no. 23, pp. 23416-23440, 2022.
[5] Y. Cui, F. Liu, X. Jing, and J. Mu, "Integrating Sensing and Communications for Ubiquitous IoT: Applications, Trends, and Challenges," *IEEE Network,* vol. 35, no. 5, pp. 158-167, 2021.
[6] D. K. P. Tan, J. He, Y. Li, A. Bayesteh, Y. Chen, P. Zhu, and W. Tong, "Integrated Sensing and Communication in 6G: Motivations, Use Cases, Requirements, Challenges and Future Directions." pp. 1-6.
[7] Z. Wei, F. Liu, C. Masouros, N. Su, and A. P. Petropulu, "Toward Multi-Functional 6G Wireless Networks: Integrating Sensing, Communication, and Security," *IEEE Communications Magazine,* vol. 60, no. 4, pp. 65-71, 2022.
[8] Y. Cui, X. Jing, and J. Mu, "Integrated Sensing and Communications Via 5G NR Waveform: Performance Analysis." pp. 8747-8751.
[9] Z. Xiao, and Y. Zeng, "Waveform Design and Performance Analysis for Full-Duplex Integrated Sensing and Communication," *IEEE Journal on Selected Areas in Communications,* vol. 40, no. 6, pp. 1823-1837, 2022.
[10] Y. Wu, F. Lemic, C. Han, and Z. Chen, "Sensing Integrated DFT-Spread OFDM Waveform and Deep Learning-Powered Receiver Design for Terahertz Integrated Sensing and Communication Systems," *IEEE Transactions on Communications,* vol. 71, no. 1, pp. 595-610, 2023.
[11] X. Tian, and Z. Song, "On radar and communication integrated system using OFDM signal." pp. 0318-0323.
[12] D. H. N. Nguyen, and R. W. Heath, "Delay and Doppler processing for multi-target detection with IEEE 802.11 OFDM signaling." pp. 3414-3418.
[13] M. Braun, C. Sturm, and F. K. Jondral, "Maximum likelihood speed and distance estimation for OFDM radar." pp. 256-261.
[14] P. Kumari, J. Choi, N. González-Prelcic, and R. W. Heath, "IEEE 802.11ad-Based Radar: An Approach to Joint Vehicular Communication-Radar System," *IEEE Transactions on Vehicular Technology,* vol. 67, no. 4, pp. 3012-3027, 2018.
[15] L. Ma, C. Pan, Q. Wang, M. Lou, Y. Wang, and T. Jiang, "A Downlink Pilot Based Signal Processing Method for Integrated Sensing and Communication Towards 6G." pp. 1-5.
[16] C. Zhang, Z. Zhou, H. Wang, and Y. Zeng, "Integrated super-resolution sensing and communication with 5G NR waveform: Signal processing with uneven CPs and experiments." pp. 681-688.
[17] E. Dahlman, S. Parkvall, and J. Skold, *5G NR: The next generation wireless access technology*: Academic Press, 2020.
[18] Z. Wei, Y. Wang, L. Ma, S. Yang, Z. Feng, C. Pan, Q. Zhang, Y. Wang, H. Wu, and P. Zhang, "5G PRS-Based Sensing: A Sensing Reference Signal Approach for Joint Sensing and Communication System," *IEEE Transactions on Vehicular Technology,* vol. 72, no. 3, pp. 3250-3263, 2023.
[19] P. K. Rai, A. Kumar, M. Z. A. Khan, and L. R. Cenkeramaddi, "LTE-based passive radars and applications: a review," *International Journal of Remote Sensing,* vol. 42, no. 19, pp. 7489-7518, 2021.
[20] A. Evers, and J. A. Jackson, "Analysis of an LTE waveform for radar applications." pp. 0200-0205.
[21] Q. Zhao, A. Tang, and X. Wang, "Reference Signal Design and Power Optimization for Energy-Efficient 5G V2X Integrated Sensing and Communications," *IEEE Transactions on Green Communications and Networking,* vol. 7, no. 1, pp. 379-392, 2023.
[22] M. A. Richards, *Fundamentals of radar signal processing*: Mcgraw-hill New York, 2005.
[23] M. Braun, M. Fuhr, and F. K. Jondral, "Spectral estimation-based OFDM radar algorithms for IEEE 802.11 a Signals." pp. 1-5.
[24] S. Jardak, S. Ahmed, and M. S. Alouini, "Low Complexity Moving Target Parameter Estimation for MIMO Radar Using 2D-FFT," *IEEE Transactions on Signal Processing,* vol. 65, no. 18, pp. 4745-4755, 2017.
[25] R. Zhang, S. Tsai, T.-H. Chou, and J. Ren, "Comb Reference Signal Pattern Design for Integrated Communication and Sensing," *arXiv preprint arXiv:2401.09648,* 2024.
[26] F. Belfiori, W. v. Rossum, and P. Hoogeboom, "Application of 2D MUSIC algorithm to range-azimuth FMCW radar data." pp. 242-245.
[27] R. Xie, D. Hu, K. Luo, and T. Jiang, "Performance Analysis of Joint Range-Velocity Estimator With 2D-MUSIC in OFDM Radar," *IEEE Transactions on Signal Processing,* vol. 69, pp. 4787-4800, 2021.
[28] W. Roberts, P. Stoica, J. Li, T. Yardibi, and F. A. Sadjadi, "Iterative Adaptive Approaches to MIMO Radar Imaging," *IEEE Journal of Selected Topics in Signal Processing,* vol. 4, no. 1, pp. 5-20, 2010.
[29] Y. Zhang, Y. Zhang, W. Li, Y. Huang, and J. Yang, "Super-Resolution Surface Mapping for Scanning Radar: Inverse Filtering Based on the Fast Iterative Adaptive Approach," *IEEE Transactions on Geoscience and Remote Sensing,* vol. 56, no. 1, pp. 127-144, 2018.
[30] M. H. Hayes, *Statistical digital signal processing and modeling*: John Wiley & Sons, 1996.

[31] T. Yardibi, J. Li, P. Stoica, M. Xue, and A. B. Baggeroer, "Source Localization and Sensing: A Nonparametric Iterative Adaptive Approach Based on Weighted Least Squares," *IEEE Transactions on Aerospace and Electronic Systems,* vol. 46, no. 1, pp. 425-443, 2010.
[32] Y. Wang, X. Huang, and R. Cao, "Novel Method of ISAR Cross-Range Scaling for Slowly Rotating Targets Based on the Iterative Adaptive Approach and Discrete Polynomial-Phase Transform," *IEEE Sensors Journal,* vol. 19, no. 13, pp. 4898-4906, 2019.
[33] G. O. Glentis, and A. Jakobsson, "Efficient Implementation of Iterative Adaptive Approach Spectral Estimation Techniques," *IEEE Transactions on Signal Processing,* vol. 59, no. 9, pp. 4154-4167, 2011.
[34] Y. Li, S. Chang, Z. Liu, W. Ren, and Q. Liu, "Range ambiguity suppression under high-resolution estimation using the MUSIC-AP algorithm for pulse-Doppler radar," *Signal Processing,* vol. 214, pp. 109237, 2024.
[35] J. B. Correll, J. K. Beard, and C. N. Swanson, "Costas Array Waveforms for Closely Spaced Target Detection," *IEEE Transactions on Aerospace and Electronic Systems,* vol. 56, no. 2, pp. 1045-1076, 2020.
[36] S. W. Golomb, and H. Taylor, "Constructions and properties of Costas arrays," *Proceedings of the IEEE,* vol. 72, no. 9, pp. 1143-1163, 1984.

**Rui Zhang** (Member, IEEE) is currently an Assistant professor at Electrical Engineering department, State University of New York at Buffalo, Buffalo NY USA. She received a Ph.D. degree at the School of Electrical and Computer Engineering, Georgia Institute of Technology, Atlanta GA USA in 2022. In 2017, she received her B.S. degree in electrical engineering and B.A. degree in economics both from Peking University, Beijing, China. From 2022 to 2023, she was a staff engineer at Communication System Design of MediaTek USA Inc. in San Diego, CA. She is the recipient of the 2022 Marconi Young scholar award and associate technical editor of IEEE Communication Magazine. Her research interests are wireless and optical communications, and signal processing.

**Shawn Tsai**. Shawn Tsai received M.S.E.E. and Ph.D. degrees from Purdue University, West Lafayette, IN, in 1995 and 2000, respectively. From 2000 to 2020, he was a system engineer in Ericsson, Huawei, and Qualcomm, contributing to research and development of wireless system standardization, design, implementation, and commercialization. He is currently an Engineering Director at Communication System Design of MediaTek USA Inc. in San Diego, CA. His interests include communications theory, signal processing, radio resource management, and air interface protocols.

**Tzu-Han Chou** is a systems engineer at MediaTek Inc., USA. Before joining MediaTek, he worked at Qualcomm Inc. in San Diego, CA, as well as the Industrial Technology Research Institute (ITRI) and Sunplus Technology in Hsinchu, Taiwan. He earned his Ph.D. in Electrical and Computer Engineering from the University of Wisconsin, Madison, in 2011, and his B.S. in Electrical Engineering and M.S. in Communication Engineering from National Taiwan University. Specializing in wireless communication, information theory, and signal processing, he brings several years of experience in the telecommunications industry. Throughout his career, he has made significant contributions to baseband development and 3GPP standardization.

**Jiaying Ren** received the B.S. degree in applied physics from the University of Science and Technology of China, Hefei, China, in 2016, and the Ph. D degree in Electrical Engineering from the University of Florida, Gainesville, FL, USA, in 2022. She is currently a Staff Engineer with the Advanced Communication Technology, MediaTek, San Diego, CA, USA. Her research interests include spectral estimation, statistical signal processing, and their applications.

**Wenze Qu** received the B.S. degree in Communication Engineering from Shandong University, Shandong, China, in 2006, and the M.S. degree in Microelectronics and Solid-State Electronics from Institute of Microelectronics, Chinese Academy of Sciences, Beijing, China, in 2009. Currently, he is a Research Engineer in MediaTek (Beijing) Inc. His current research interest is integrated sensing and communication in 5G-A/6G.

**Oliver Sun** is currently Assistant General Manager of Communication System Design headquarter at MediaTek, Inc., leading advanced wireless research and standardization, commercial system design, proof of concept prototyping and strategic technology roadmap. He received his M.S. degree in communication engineering from National Taiwan University in 2001. Oliver received top innovation and contributor awards in MediaTek. He contributed cellular system algorithm and architecture design and led the teams to develop modem generations from 2G to 5G. Starting from 2019, Oliver has been driving MediaTek standard research in 3GPP, technical ecosystem partnership, disruptive technology roadmap pipeline, and advanced R&D for 6G innovation.